\documentclass[sigconf]{acmart}
\AtBeginDocument{%
  \providecommand\BibTeX{{%
    \normalfont B\kern-0.5em{\scshape i\kern-0.25em b}\kern-0.8em\TeX}}}

\setcopyright{acmcopyright}
\copyrightyear{2024}
\acmYear{2024}
\acmDOI{XXXXXXX.XXXXXXX}

\acmConference[Conference SIGMOD '24]{In Proceedings
of the 2024 International Conference on Management of Data}{June 09--15,
  2024}{Santiago, Chile}
\acmPrice{15.00}
\acmISBN{978-1-4503-XXXX-X/18/06}

\usepackage{bbding}

\usepackage{multirow}

\usepackage{booktabs}
\usepackage{textcomp}
\usepackage{xcolor} % 用于设置颜色
\usepackage{subfigure}
\usepackage{algorithmicx}
\usepackage{algorithm}
\usepackage{algpseudocode}
\usepackage{diagbox}
\usepackage{makecell}%为了表格换行

\usepackage{graphics}
\usepackage{graphicx}
\usepackage{twemojis}

\usepackage{svg}
\usepackage{xspace}

\usepackage{balance}
\usepackage{tikz} % 用于绘制图形和圆圈

\definecolor{mycolororange}{RGB}{255,192,0}
\definecolor{mycolorgreen}{RGB}{0,176,80}

\newcommand{\coloredcircled}[2][mycolororange]{\tikz[baseline=(char.base)]{\node[shape=circle,draw,inner sep=1pt,fill=#1,text=black] (char) {#2};}} %带颜色带圈的数字实现

\newcommand{\coloredcircledgreen}[2][mycolorgreen]{\tikz[baseline=(char.base)]{\node[shape=circle,draw,inner sep=1pt,fill=#1,text=black] (char) {#2};}} %带颜色带圈的数字实现

\newcommand{\idxname}{SALI\xspace}

\begin{document}

%\title{\idxname: A Scalable Adaptive Learned Index based on Probability Model}
\title{\idxname: A Scalable Adaptive Learned Index Framework based on Probability Models}


% \affiliation{%

\author{Jiake Ge$^{\spadesuit}$$^{\heartsuit}$, Huanchen Zhang$^{\diamondsuit}$$^{\mathsection}$, Boyu Shi$^{\clubsuit}$$^{\heartsuit}$, Yuanhui Luo$^{\spadesuit}$$^{\heartsuit}$, Yunda Guo$^{\spadesuit}$$^{\heartsuit}$, Yunpeng Chai$^{\clubsuit}$$^{\heartsuit}$$^\ast$, Yuxing Chen$^+$, Anqun Pan$^+$}
\authornote{Yunpeng Chai is the corresponding author.}
\affiliation{%
  \institution{$^{\spadesuit}$Key Laboratory of Data Engineering and Knowledge Engineering, MOE, China}
  \city{}
  \country{}
}
\affiliation{%
  \institution{$^{\clubsuit}$Engineering Research Center of Database and Business Intelligence, MOE, China}
  \city{}
  \country{}
}
\affiliation{%
  \institution{$^{\heartsuit}$School of Information, Renmin University of China}
  \city{}
  \country{}
}
\affiliation{%
  \institution{$^{\mathsection}$China and Shanghai Qi Zhi Institute}
  \city{}
  \country{}
}
\affiliation{%
  \institution{$^{\diamondsuit}$Tsinghua University}
  \city{}
  \country{}
}
\affiliation{%
  \institution{$^+$Tencent Inc.}
  \city{}
  \country{}
}
\affiliation{%
  \institution{gejiake@ruc.edu.cn, \hspace{0.5em}huanchen@tsinghua.edu.cn, \hspace{0.5em}shiboyu5687@ruc.edu.cn, \hspace{0.5em}losk@ruc.edu.cn, \hspace{0.5em}guoyunda@ruc.edu.cn, \hspace{0.5em}ypchai@ruc.edu.cn, \hspace{0.5em}axingguchen@tencent.com, \hspace{0.5em}aaronpan@tencent.com}
  \city{}
  \country{}
}

\renewcommand{\shortauthors}{Jiake Ge and Huanchen Zhang, et al.}

%
% The abstract is a short summary of the work to be presented in the
% article.
\begin{abstract}

The growth in data storage capacity and the increasing demands for high performance have created several challenges for concurrent indexing structures. 
% Designing scalable yet efficient concurrent index structures has become a crucial issue.
One promising solution is learned indexes, which use a learning-based approach to fit the distribution of stored data and predictively locate target keys, significantly improving lookup performance. 
Despite their advantages, prevailing learned indexes exhibit constraints and encounter issues of scalability on multi-core data storage.

This paper introduces \idxname, the \textbf{S}calable \textbf{A}daptive \textbf{L}earned \textbf{I}ndex framework, which incorporates two strategies aimed at achieving high scalability, improving efficiency, and enhancing the robustness of the learned index.
Firstly, a set of node-evolving strategies is defined to enable the learned index to adapt to various workload skews and enhance its concurrency performance in such scenarios.
Secondly, a lightweight strategy is proposed to maintain statistical information within the learned index, with the goal of further improving the scalability of the index.
Furthermore, to validate their effectiveness, \idxname applied the two strategies mentioned above to the learned index structure that utilizes fine-grained write locks, known as LIPP.
The experimental results have demonstrated that \idxname significantly enhances the insertion throughput with 64 threads by an average of $2.04\times$ compared to the second-best learned index. Furthermore, \idxname accomplishes a lookup throughput similar to that of LIPP+.

\end{abstract}


% 若正式见刊，需要去掉注释！！！！！！！！！！！！！！！！！！！！！！！！！！！！！！！！！！！！！！！！！！！！！！！


%到这里！！！！！！！！！！！！！！！！！！！！！！！！！！！！！！！！！！！！！！！！！！！！！！！！！！！！！


\vspace{-1em}

%% A "teaser" image appears between the author and the affiliation

\received{15 April 2023}
\received[revised]{20 July 2023}
\received[accepted]{23 August 2023}

\settopmatter{printfolios=true}

\maketitle

\section{Introduction}
\label{sec:intro}

With the exponential growth of the data volume today,
efficient indexing data structures are crucial for a big data system
to support timely information retrieval.
To improve the performance and memory efficiency of traditional tree-based indexes,
Kraska et al. introduced a learned index, called the Recursive Model Indexes (RMI)
that uses machine learning models to replace the internal nodes of a B+tree~\cite{kraska2018case, levandoski2013bw}.
An outstanding problem of the original RMI is that it is static:
inserting or updating a key in the index requires a significant portion of
the data structure to rebuild, thus limiting the use cases of the learned index.

Previous work has proposed two strategies to address the updatability issue of learned indexes.
The first (i.e., the buffer-based strategy) is to accommodate new entries in separate
insert buffers first to amortize the index reconstruction cost.
XIndex~\cite{tang2020xindex} and FINEdex~\cite{li2021finedex} fall into this category.
The other strategy (i.e., the model-based strategy) adopted by
ALEX~\cite{ding2020alex} and LIPP~\cite{wu2021updatable}
is to reserve slot gaps within nodes to handle new entries with an in-place insertion.
Upon an insert collision (i.e., the mapped slot is already occupied),
ALEX reorganizes the node by shifting the existing entries,
while LIPP utilizes a chaining scheme, creating a new node for the corresponding slot to transform the last-mile search problem into a sub-tree traversal problem.

\begin{figure}
  \vspace{-1em}
  \centering
  \includegraphics[width=0.8\linewidth]{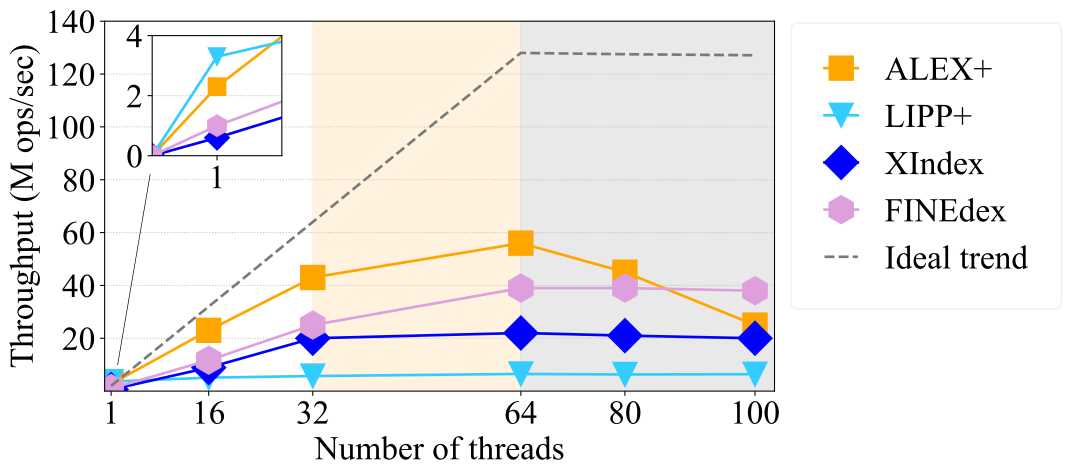}
  \vspace{-1.3em}
  \caption{Write-only performance of state-of-the-art learned indexes on the FACE dataset~\cite{facebook}. The evaluation is conducted on a two-socket machine with two 16-core CPUs.}
  \vspace{-2em}
  \label{fig:figure 0}
\end{figure}

We found, however, that none of the above index designs scale at a high concurrency.
We performed an experiment where we insert 200 million random integer keys into
a learned index, with a varying number of threads each time.
\autoref{fig:figure 0} shows the results.
Note that the number of threads in the grey area of the figure is larger than
the number of hardware threads of the machine.
This is common in practice as a database/key-value server typically handles
a large number of user connections simultaneously.

As shown in \autoref{fig:figure 0},
indexes with a buffer-based strategy (i.e., XIndex and FINEdex) exhibit
inferior base performance and worse scalability compared to those
with a model-based strategy (i.e., ALEX+ and LIPP+\footnote{ALEX+ and LIPP+ are concurrent implementations of ALEX and LIPP, respectively~\cite{gre}.}).
This shows that \textbf{a larger margin of prediction errors prevents scaling}
because the concurrent ``last mile'' searches saturate the memory bandwidth quickly
which becomes the system's bottleneck~\cite{gre}.

The problem is solved in LIPP+ where each position prediction is accurate (i.e., no ``last mile'' search).
However, LIPP+ requires maintaining statistics, such as access counts and collision counts, in each node
to trigger node retraining to prevent performance degradation.
These \textbf{per-node counters create high contention} among threads and cause severe
cacheline ping-pong~\cite{gre}.
The model-based strategy in ALEX+ requires shifting existing entries upon an insert collision.
Therefore, ALEX+ must acquire \textbf{coarse-grained write locks} for this operation.
As the number of threads increases, more and more threads are blocked, waiting for those
exclusive locks.

% In this paper, we propose \idxname,
% the \textbf{S}calable \textbf{A}daptive \textbf{L}earned \textbf{I}ndex
% based on probability models to solve the scalability issues in existing solutions.
% \idxname adopts the model-based insertion strategy with chaining to minimize the
% searching overhead and to save memory bandwidth.
% This design also allows \idxname to use fine-grained (i.e., per slot) locks
% to handle concurrent inserts/updates.
% To solve the scalability bottleneck of maintaining per-node statistics,
% we developed lightweight probability models that can trigger node retraining
% and other structural evolving operations in \idxname with accurate timing
% (as if the timing were determined by accurate statistics).
% In addition, we developed a set of node-evolving strategies, including
% expanding an insert-heavy node to contain more gaps, flattening the tree
% structure for frequently-accessed nodes, and compacting the rarely-touched
% nodes to save memory.
% \idxname applies these node-evolving strategies adaptively according to
% the probability models so that it can self-adjust to changing workloads.

In this paper, we propose \idxname,
the \textbf{S}calable \textbf{A}daptive \textbf{L}earned \textbf{I}ndex framework based on
probability models to solve the scalability issues in existing solutions.
To solve the scalability bottleneck of maintaining per-node statistics,
we developed lightweight probability models that can trigger node retraining
and other structural evolving operations in \idxname with accurate timing
(as if the timing were determined by accurate statistics).
In addition, we developed a set of node-evolving strategies, including
expanding an insert-heavy node to contain more gaps, flattening the tree
structure for frequently-accessed nodes, and compacting the rarely-touched
nodes to save memory.
\idxname applies these node-evolving strategies adaptively according to
the probability models so that it can self-adjust to changing workloads while maintaining excellent scalability.
Finally, \idxname adopts the learned index structure that utilizes fine-grained write locks, i.e., LIPP+, to validate the effectiveness of the aforementioned two strategies.
Note that the lightweight probability models and node-evolving strategies are highly versatile and can be applied to various index scenarios, as detailed in Section \ref{discussion}.
% This structure can minimize the searching overhead and save memory bandwidth and also allows \idxname to use fine-grained (i.e., per slot) locks to handle concurrent inserts/updates.

Our microbenchmark with real-world data sets shows that
\idxname improves the insertion throughput with 64 threads by 
$2.04\times$ on average
compared to the second-best learned index, i.e., ALEX+,
while achieving a lookup throughput comparable to LIPP+.

We make three primary contributions in this paper.
Firstly, we proposed SALI, a high-concurrency learned index framework designed to improve the scalability of learned indexes.
Secondly, we defined a set of node-evolving strategies in addition to model retraining
to allow the learned index to self-adapt to different workload skews.
Thirdly, we replaced the per-node statistics in existing learned indexes with 
lightweight probability models to remove the scalability bottleneck of statistics
maintenance while keeping the timing accuracy of node retraining/evolving.
Finally, we proved the effectiveness of the proposed approaches by showing that
\idxname outperforms the SOTA learned indexes under high concurrency.

The rest of this paper is organized as follows.
Section~\ref{background and motivation} summarizes the basics of learned indexes and further motivates the scalability problem.
Section~\ref{the design of sali} introduces the structure of \idxname with an emphasis on the node-evolving strategies
and the probability models.
Section~\ref{evalution} presents our experimental results.
Section~\ref{discussion} discusses the generalizability of the node-evolving strategies and the probability models, as well as the limitations of SALI, followed by a related work discussion in Section~\ref{related work}
Section~\ref{conclusion} concludes the paper.
The source code of this paper has been made available at \url{https://github.com/YunWorkshop/SALI}.

% \vspace{-0.6em}
\section{Background and Motivation}
\label{background and motivation}

\begin{figure}
\vspace{-1.8em}
  \centering
  \includegraphics[width=\linewidth]{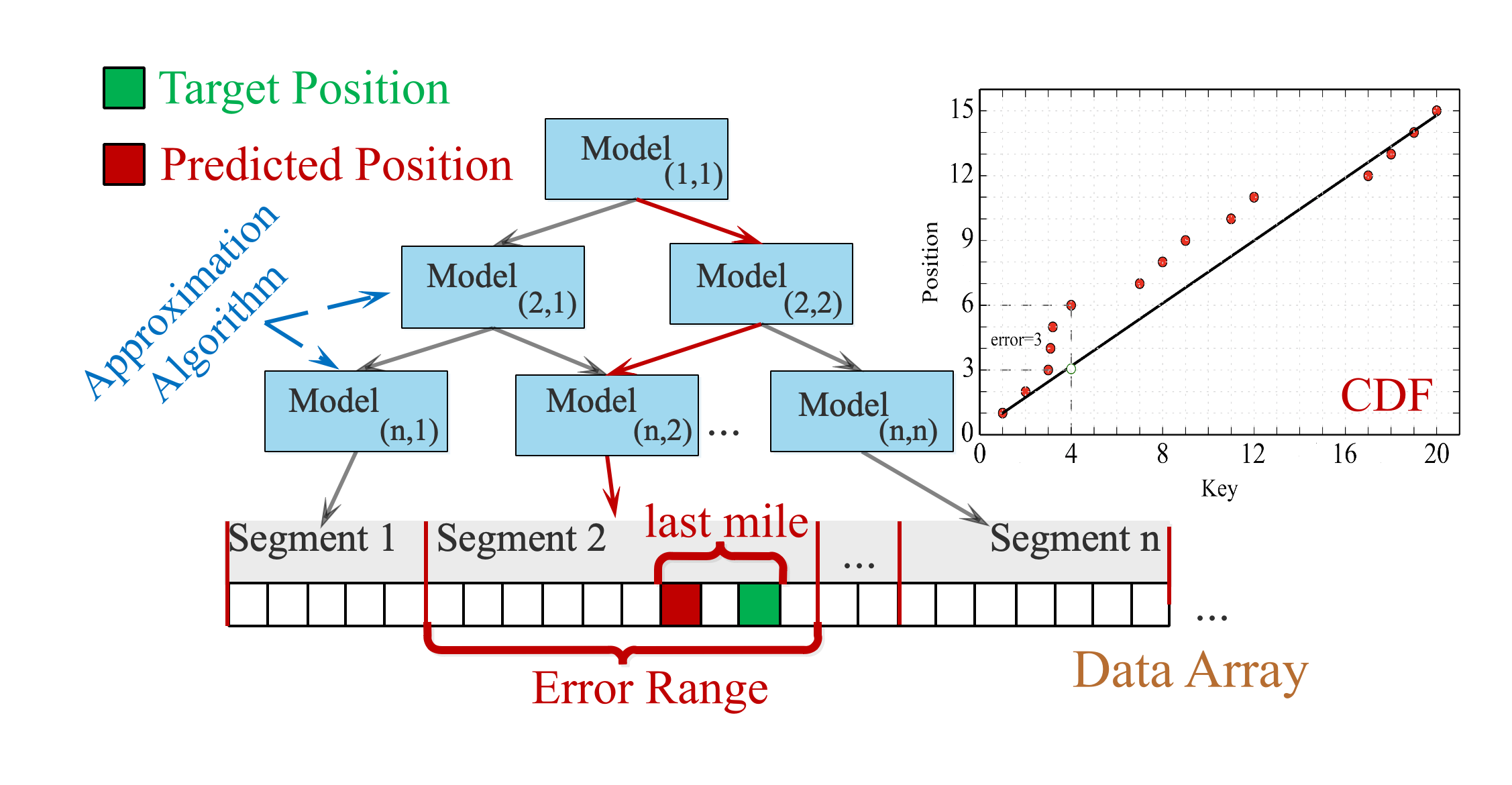}
  \vspace{-3.6em}
  \caption{The scheme of the learned index.}
  \vspace{-2.2em}
  \label{fig:figure 1}
\end{figure}

\subsection{The principle of learned indexes}
\label{The principle of learned indexes}

The core concept of the learned index is to employ a set of learning models to estimate the cumulative distribution function (CDF) of the stored data~\cite{marcus2020benchmarking}, allowing for the prediction of the data's storage location, as depicted in the CDF diagram on \autoref{fig:figure 1}.

\autoref{fig:figure 1} shows the scheme for the learned index structure. 
Each node, or only the lowest leaf nodes, stores the slope and intercept of the linear function~\cite{ferragina2020pgm,ding2020alex,kraska2018case,tang2020xindex,li2021finedex,wu2021updatable, galakatos2019fiting}. 
Each segment corresponds to a linear model, which is responsible for the approximate position of the target key. 
The index segments correspond to linear models that estimate the target key's position, eliminating the need for multiple indirect search operations in traditional tree-based indexes. 
This approach has the potential to improve indexed lookup performance significantly.

% \vspace{-0.5em}
\subsection{Scalable Evaluation of Learned Index Structures} \ \
\label{A concurrency-friendly learned index structure}
\vspace{-1em}

Currently, learned indexes demonstrate good performance in single-threaded environments. 
However, their scalability remains limited~\cite{gre}. 
In this part, our objective is to conduct a thorough investigation into the factors that contribute to the concurrent performance bottlenecks in existing learned indexes. 
To achieve this, we begin by introducing the insertion strategies employed in learned indexes, along with their corresponding index structures, as these design choices significantly influence the concurrent performance of the indexes. 
Additionally, we conduct a comprehensive evaluation of the index structures and identify their limitations in terms of scalability.

% \vspace{-0.5em}
\subsubsection{The insertion strategies of learned indexes}

In a concurrent scenario, the blocking of index operations is primarily due to the insertion of new keys. 
Understanding the current insertion strategies is essential for enhancing index scalability. 
Thus, we present the insertion strategies as follows.

\begin{figure}
\vspace{-2.5em}
  \centering
  \includegraphics[width=\linewidth]{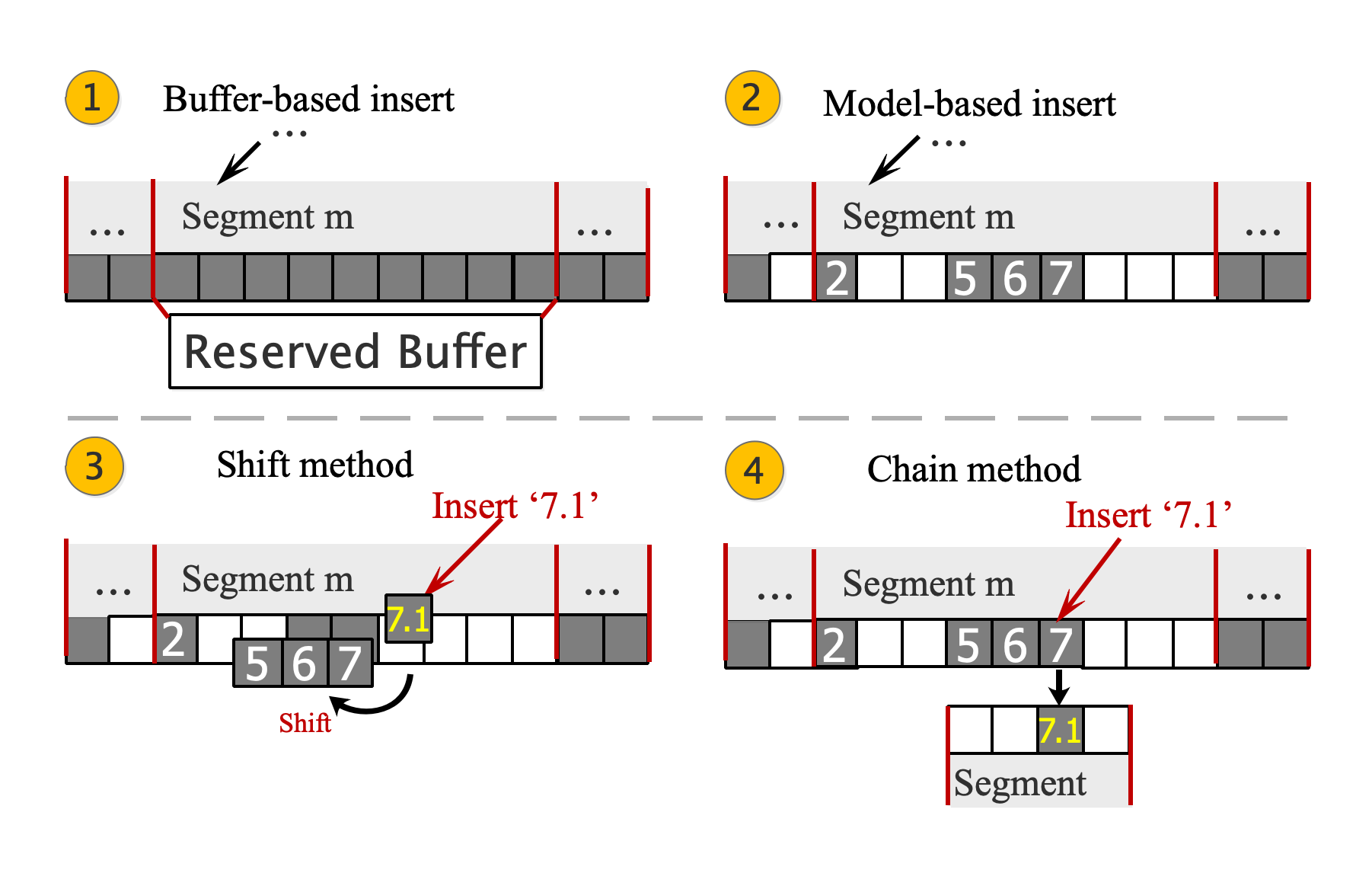}
  \vspace{-3.5em}
  \caption{The updatable strategies in learned indexes.}
  \label{fig:figure 2}
  \vspace{-2em}
\end{figure}

\textbf{Strategy 1:}
Scholars try to design a buffer-based insert strategy, i.e., off-site insertion, on learned indexes to implement insert operations~\cite{galakatos2019fiting, ferragina2020pgm, tang2020xindex, li2021finedex}.
As shown in \autoref{fig:figure 2}, the core idea of the buffer-based strategy is to create a buffer structure for insertion~\cite{galakatos2019fiting, tang2020xindex, li2021finedex, ferragina2020pgm}. 
When the buffer is full, its keys must be merged with those in the upper segment and transformed into a new linear model. 
Furthermore, this structure suffers from significant errors due to the intensive storage of keys (no gap)~\cite{ge2023learnedindexevaluation}.

\textbf{Strategy 2:} 
The core idea of the model-based insert strategy, i.e., in-place insertion,  involves reserving gaps in the nodes~\cite{ding2020alex,wu2021updatable}. 
If the linear model predicts that the target position of the inserted key is a gap, it is directly inserted.
However, if the slot with the key already exists, there are two existing conflict resolution strategies:

•	\textit{Solution 1:} 
One potential solution for resolving conflicts is to adopt the ``shift'' method~\cite{ding2020alex}. 
As shown in \autoref{fig:figure 2}, this method involves shifting the existing conflicted key and its adjacent keys to the nearest gap by one slot, allowing the target key to be inserted into the target position. However, the process of key movement can introduce errors.

•	\textit{Solution 2:}  
Another solution to resolve conflicts is to adopt the ``chain'' method~\cite{wu2021updatable}. 
As shown in \autoref{fig:figure 2}, if a key already exists at the target position of the newly inserted key, a new node is created downward to accommodate the conflicting key.
This conflict resolution approach does not involve moving any data, thus avoiding potential lookup errors (precise lookup).

% \vspace{-0.5em}
\subsubsection{In-depth analysis of these strategies}

This part will provide an in-depth analysis of the index structures corresponding to the insertion strategies mentioned.
Building upon the GRE~\cite{gre}, we further performed an in-depth experimental analysis of existing learned indexes and identified the scalability problems in their designs.
Our objective is not to compare them with each other but to highlight the scalability bottleneck.

\begin{figure}[t]
\vspace{-1em}
  \centering
  \includegraphics[width=\linewidth]{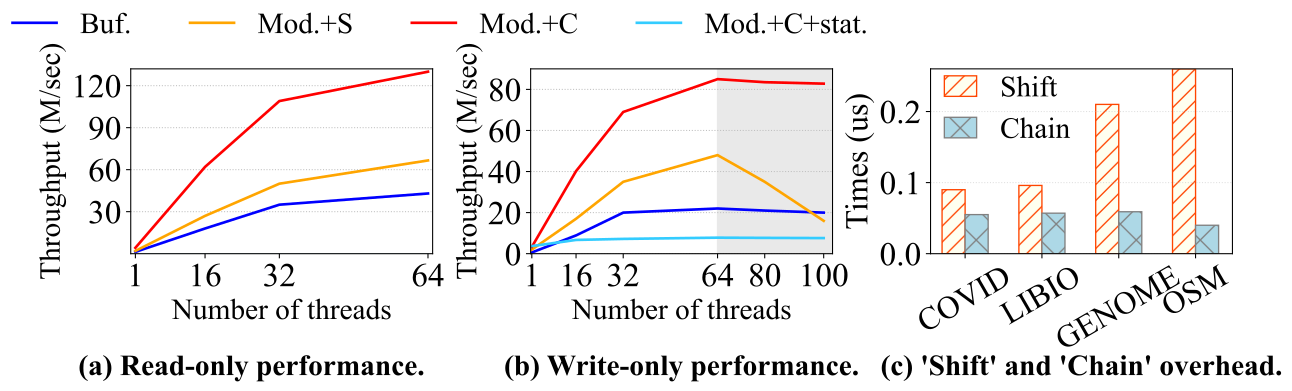}
  \vspace{-2.4em}
  \caption{The in-depth evaluation of performance on the COVID dataset~\cite{gre}, where the workload follows the uniform distribution. The evaluation is conducted on a two-socket machine with two 16-core CPUs.}
  \label{fig:figure 6}
  \vspace{-2em}
\end{figure}

\autoref{fig:figure 6} illustrates the performance for the three structures. 
The buffer-based structure is denoted by $buf.$ (i.e., the structure of XIndex), the \textit{\textbf{Mod}}el-based strategy with the \textit{\textbf{S}}hift method is denoted by $Mod.+S$ (i.e., the structure of ALEX), the \textit{\textbf{Mod}}el-based strategy with the \textit{\textbf{C}}hain method is denoted by $Mod.+C$ (i.e., the structure of LIPP).
The notation $Mod.+C+stat.$ is used to represent the $Mod.+C$ approach along with the maintenance of \textit{\textbf{stat}}istics to track index deterioration. 
Note that except for $Mod.+C+stat.$, we disabled the statistics maintenance and local adjustment functions with the purpose of analyzing the impact of the structures themselves on index concurrency performance.

\autoref{fig:figure 6}(a) displays that both $buf.$ and $Mod.+S$ exhibit poor scalability due to lookup errors, which can impact both the lookup and insert performance.
Note that, the average search error of $buf.$ is higher compared to $Mod.+S$~\cite{ge2023learnedindexevaluation}.

\textbf{Observation 1: Improved concurrency performance can be achieved through the utilization of precise lookups.
The insertion strategy employed by \textit{Mod.+C} guarantees error-free generation even during the insertion process.}

\autoref{fig:figure 6}(b) depicts the insertion performance for the three structures. 
The poor performance of $buf.$ is attributed to a lookup error and an off-site insertion method~\cite{ge2023learnedindexevaluation}. 
$Mod.+S$ has three reasons for its poor performance: 
(1) Severe lookup errors; 
(2) Significant write amplification and frequent ``last mile'' lookup lead to exhaustion of memory bandwidth and affect concurrency performance~\cite{gre}.
This amplification occurs because the moving key and the inserted target key need to be written into the node together;
(3) During insertion, coarse-grained locks can easily cause thread blocking.
The shift method leads to significant correlations between keys in the entire node, necessitating the locking of the entire node during insertion to ensure accuracy.
Moreover, the thread blocking issue is more pronounced in the gray area in \autoref{fig:figure 6}(b), where threads frequently access nodes locked by coarse-grained locks and release CPU time slices, resulting in invalid operations and exacerbating performance degradation.
In fact, the coarse-grained locks have already caused a slowdown in concurrent performance growth under 64 threads, which may not have been apparent due to the influence of other scalability factors, e.g., lookup error.

\textbf{Observation 2: 
\textit{Mod.+C} facilitates fine-grained locks, i.e., one slot, no search errors, and in-place insertion, which enables its good insertion scalability.}

The scalability of $Mod.+C+stat.$ is severely impacted, depicted in \autoref{fig:figure 6}(b), primarily due to the high level of contention and cache-line ping-pong that arises from maintaining statistics in a concurrent scenario, which is consistent with the findings of Wongkham et al.~\cite{gre}.
\autoref{fig:figure 6}(c) shows that $Mod.+C$ saves 0.8x-5x the average time compared to $Mod.+S$ with different datasets.

\textbf{Observation 3: 
Maintaining statistics within \textit{Mod.+C} renders the index non-scalable. 
Note that the scalability issue is not attributed to the structure of \textit{Mod.+C}, but rather to the absence of a lightweight statistical approach.}

\textbf{Observation 4: 
The chain method exhibits significantly shorter operation times compared to the shift method.}

% \subsubsection{Lessons learned}
% Based on the above four observations and our philosophy of prioritizing concurrency as a first citizen, we determined that \idxname is designed using $Mod.+C$ as the primary index structure.
% Note that we have developed a lightweight methodology to maintain statistics, which is discussed in detail in Section~\ref{Probability Model}.
% Additionally, we anticipate that $Mod.+C$ will serve as a fundamental strategy for designing updatable learned indexes in future research. 
% This exceptional performance strategy will pave the way for numerous learned indexes to emerge.
% This structure is expected to contribute significantly to the development of future learned indexes in diverse dimensions.

% \vspace{-0.5em}
\subsection{The Scalable Learned Index Requirements}
% \vspace{-0.2em}
\label{The Scalable Learned Index Requirements}

In consideration of scalability, we have further summarized the potential limitations of several SOTA learned indexes in Table \ref{tab:table 1}, taking into account factors such as prediction accuracy ($No\ err$), insert strategy ($In$-$place$), lock granularity ($Fine\ lock$), lightweight statistics support ($l$-$w\ stat.$), etc. 
A checkmark denotes support for the given factor, while a cross sign indicates the lack of support.

We believe that \textbf{designing a learned index requires prioritizing concurrency control and robustness as first-class considerations, adopting a holistic approach to ensure consistency in design choices.}
Therefore, considering the challenges associated with learned indexes, we propose that a more scalable learned index should simultaneously address the following dimensions:

% \vspace{-0.5em}
\subsubsection{Efficient concurrency}\ \

1) Maintaining statistics should barely impact scalability.
To enable efficient insertion performance, updatable learned indexes must track statistical information that reflects the degradation of the index structure over time due to new insertions. 
This information is crucial for performing necessary retraining operations. 
% Commonly used statistics in learned indexes include the amount of data inserted into a node, the number of insertion conflicts, and the remaining reserved gaps. 
However, maintaining these cumulative statistics jointly by the insertion thread can potentially lead to blocking, becoming a scalability bottleneck for some state-of-the-art learned indexes~\cite{gre}. 
Therefore, there is an urgent need to develop a lightweight methodology to maintain statistics.

2) Designing effective index structures for concurrent scenarios.
In concurrent scenarios, insertion performance in learned indexes can be hampered by blocking that often arises when multiple insertion threads work together to uphold key consistency in a single local structure, particularly under skewed workloads. 
To mitigate this issue, reducing the manipulation of already-stored keys during the insertion of new ones can help minimize lock granularity and lower the risk of thread blocking~\cite{li2021finedex}.

% \vspace{-0.5em}
\subsubsection{Adaptive ability} \ \

The learned index exhibits suboptimal performance under skewed insertion workloads compared to uniform workloads. 
The lack of workload-aware adaptive adjustment capability is the primary cause of this deficiency. 
Therefore, it is critical for a learned index to possess the adaptive capacity to guarantee its robustness in concurrent scenarios.
In addition, the learned index lacks an optimization adjustment strategy for the lookup operation, which hinders its ability to maximize lookup efficiency in concurrent scenarios. 
Furthermore, learned indexes have yet to fully capitalize on opportunities for significantly reducing index space costs under skewed workloads~\cite{anneser2022hybird}.

\vspace{-0.5em}
\subsubsection{Low overheads of basic performance} \ \

1) Efficient lookup.
Achieving high lookup performance in learned indexes typically hinges on minimizing prediction errors for lookups, as substantial errors can lead to many ``last mile'' operations in a concurrent scenario. These operations consume additional memory bandwidth and negatively impact concurrent performance~\cite{gre}.

2) Efficient insert.
Adopting the model-based strategy (i.e., the in-place insertion) rather than the buffer-based strategy (i.e., the off-site insertion), can significantly enhance the insert performance of learned indexes by reserving gaps in each node (see Section~\ref{A concurrency-friendly learned index structure}).

\begin{table}% h asks to places the floating element [h]ere.
  % \vspace{-1em}
  \small
  \caption{The limited scalability of existing schemes.}
  \vspace{-1.3em}
  \label{tab:table 1}
  \begin{tabular}{*{8}{c}}
    \toprule
    \multicolumn{1}{c}{Learned} & \multicolumn{2}{c}{Basic perform.} & \multicolumn{2}{c}{Concurrency} & \multicolumn{1}{c}{Evolv.}\\
    \cmidrule(lr){2-3}\cmidrule(lr){4-5}
    % Learned & Efficient & Sustainable & Concurrency\\
    index   &    No err & In-place  &  Fine lock & l-w stat.   & ability \\
    \midrule
    % \O & 1 in 1000& For Swedish names  \scalebox{1.3}[1]{$\times$} \raisebox{0.7ex}{\scalebox{1}{$\sqrt{}$}}\\
    RMI\cite{kraska2018case}     & \scalebox{1.3}[1.3]{$\times$}     & \scalebox{1.3}[1.3]{$\times$} & \scalebox{1.3}[1.3]{$\times$} & \scalebox{1.3}[1.3]{$\times$} & \scalebox{1.3}[1.3]{$\times$}\\
    FITing\cite{galakatos2019fiting} & \scalebox{1.3}[1.3]{$\times$} & \scalebox{1.3}[1.3]{$\times$} & \scalebox{1.3}[1.3]{$\times$} & \scalebox{1.3}[1.3]{$\times$} & \scalebox{1.3}[1.3]{$\times$}\\
    PGM\cite{ferragina2020pgm}   & \scalebox{1.3}[1.3]{$\times$}     & \scalebox{1.3}[1.3]{$\times$} & \scalebox{1.3}[1.3]{$\times$} & \scalebox{1.3}[1.3]{$\times$} & \scalebox{1.3}[1.3]{$\times$} \\
    ALEX+\cite{ding2020alex}     & \scalebox{1.3}[1.3]{$\times$}     & \raisebox{0.7ex}{\scalebox{1}{$\sqrt{}$}}  & \scalebox{1.3}[1.3]{$\times$} &\raisebox{0.7ex}{\scalebox{1}{$\sqrt{}$}} & \scalebox{1.3}[1.3]{$\times$} \\
    LIPP+\cite{wu2021updatable} & \raisebox{0.7ex}{\scalebox{1}{$\sqrt{}$}} & \raisebox{0.7ex}{\scalebox{1}{$\sqrt{}$}}  & \raisebox{0.7ex}{\scalebox{1}{$\sqrt{}$}}  & \scalebox{1.3}[1.3]{$\times$} & \scalebox{1.3}[1.3]{$\times$} \\
    XIndex\cite{tang2020xindex} & \scalebox{1.3}[1.3]{$\times$}      & \scalebox{1.3}[1.3]{$\times$}  & \raisebox{0.7ex}{\scalebox{1}{$\sqrt{}$}} & \scalebox{1.3}[1.3]{$\times$} & \scalebox{1.3}[1.3]{$\times$} \\
    FINEdex\cite{li2021finedex} & \scalebox{1.3}[1.3]{$\times$}      & \scalebox{1.3}[1.3]{$\times$}  & \raisebox{0.7ex}{\scalebox{1}{$\sqrt{}$}} & \scalebox{1.3}[1.3]{$\times$} & \scalebox{1.3}[1.3]{$\times$} \\
    \textbf{\idxname} & \raisebox{0.7ex}{\scalebox{1}{\boldmath{$\sqrt{}$}}} & \raisebox{0.7ex}{\scalebox{1}{\boldmath{$\sqrt{}$}}} & \raisebox{0.7ex}{\scalebox{1}{\boldmath{$\sqrt{}$}}} & \raisebox{0.7ex}{\scalebox{1}{\boldmath{$\sqrt{}$}}} & \raisebox{0.7ex}{\scalebox{1}{\boldmath{$\sqrt{}$}}} \\
  \bottomrule
\end{tabular}
% \caption{}
\vspace{-1.8em}
\end{table}

% \subsubsection{Lessons learned}
% Based on the above four observations and our philosophy of prioritizing concurrency as a first citizen, we determined that \idxname is designed using $Mod.+C$ as the primary index structure.
% Note that we have developed a lightweight methodology to maintain statistics, which is discussed in detail in Section~\ref{Probability Model}.
% Additionally, we anticipate that $Mod.+C$ will serve as a fundamental strategy for designing updatable learned indexes in future research. 
% This exceptional performance strategy will pave the way for numerous learned indexes to emerge.
% This structure is expected to contribute significantly to the development of future learned indexes in diverse dimensions.

% \vspace{-0.5em}
\section{\idxname: A Probability-Based Evolvable Learned Index Framework}
\label{the design of sali}

% Sections 1 and 2 highlight scalability issues with currently learned indexes that create performance bottlenecks, and the critical requirements for learned indexes to achieve scalability in concurrent scenarios.

% (1) Learned index structure design leads to large insertion and lookup overhead. i.e., a coarse-grained lock structure can cause many insertion conflicts, and lookup errors can cause memory bandwidth exhaustion. 
% (2) Insertion conflicts will become more severe under skewed workloads because the high contention occurs within a smaller index structure to compound the issue. 
% (3) Existing high-contention statistical indexes deteriorate information in a way that seriously impresses index scalability. 

This section introduces the \idxname framework, which addresses the concerns and requirements discussed in Section~\ref{sec:intro} and~\pageref{background and motivation}, and facilitates the efficient scalability of learned indexes.
Specifically, Section~\ref{overview} introduces the overall architecture of \idxname, including the architecture built upon $Mod.+C$ (i.e., the structure of LIPP) (columns 2-4 of~\autoref{tab:table 1}).
In Section~\ref{Evolving strategies}, we proposed an adaptive evolving (adjustment) strategy to further improve the robustness of the learned index under skewed and uniform workloads (columns 6 of~\autoref{tab:table 1}).
Section~\ref{Probability Model} designs a probability-based lightweight method to maintain statistics of different roles at a meager cost.
This method solves the concurrency performance bottleneck problem caused by the existing high contention statistics method (columns 5 of~\autoref{tab:table 1}).

% \vspace{-0.5em}
\subsection{Overview}
\label{overview}

\subsubsection{\idxname framework}

% SALI的主要创新点是developed a lightweight methodology to maintain statistics 以及 设计了一种adaptive evolving strategy.
% 在section2.2和2.3中，我们认为可扩展性最好的是结构MOD.+C。因此我们确定SALI以$Mod.+C$作为它的的索引结构来实现和评估我们创新的两种策略。
% 注意，我们在section 5中详细讨论了 the lightweight maintain statistics methodology以及evolving strategy的普适性。

This part introduces the structure of the \idxname framework, which encompasses the introduction of a probability-based lightweight methodology for statistics maintenance and the implementation of an adaptive evolving strategy.
Due to the common occurrence of skewed workloads in real-world environments, it is advisable to apply different evolution strategies to nodes with varying degrees of read-write hotness.
These strategies can serve as alternatives to the traditional retraining method, as they are specifically designed to enhance concurrent performance and reduce the overhead of index space (Section~\ref{Evolving strategies}).
Furthermore, \idxname utilizes a probability-based lightweight method to maintain statistics while keeping the timing accuracy of node retraining/evolving without blocking insertion operations from multiple threads, unlike the traditional approach of globally maintaining statistics (Section~\ref{Probability Model}).

\begin{figure}
\centering
\vspace{-1em}
\hspace{-0.5cm}
  \centering
  \includegraphics[width=1.05\linewidth]{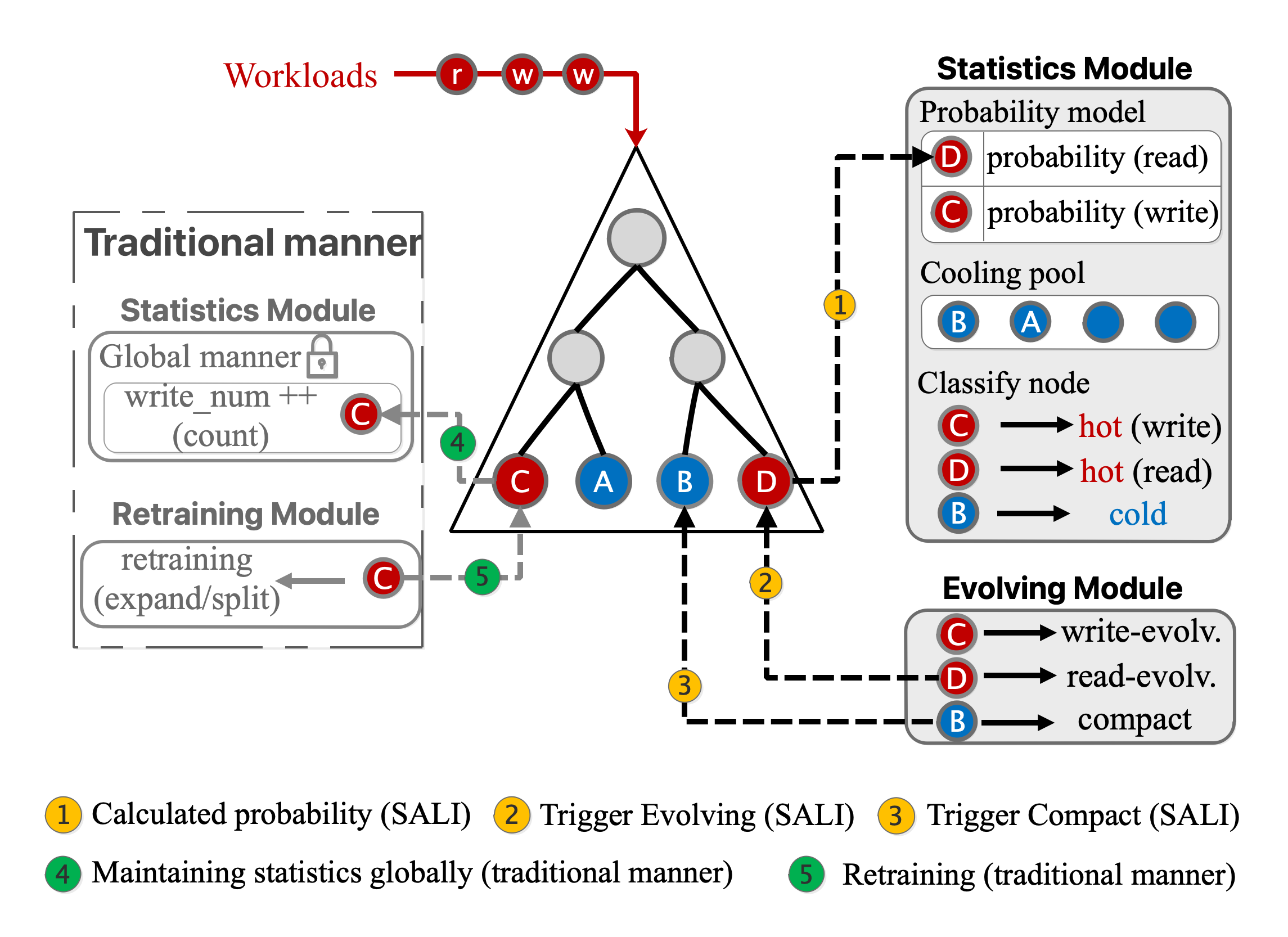}
  \vspace{-3em}
  \caption{The structure of \idxname.}
  \vspace{-2em}
  \label{fig:SALI}
\end{figure}

Specifically, the \idxname framework consists of two phases in ~\autoref{fig:SALI}. 
In the first phase, we calculate the probabilities \coloredcircled{1} for each node that requires evolving during lookup/insert operations to improve performance.
Calculating and maintaining probability models is a lightweight alternative to maintaining statistics, which does not cause high contention and thread blocking issues (see Section \ref{Probability Model}).
On the contrary, the traditional manner globally maintains statistics \coloredcircledgreen{4}, leading to high contention among threads and limiting index scalability.
In the second phase, we perform evolving operations on nodes that are classified as hot \coloredcircled{2} or cold \coloredcircled{3}. 
Note that our evolving strategy encompasses the functionality of retraining operations in the traditional manner \coloredcircledgreen{5} (refer to Section~\ref{Evolving strategies}).

\vspace{-0.5em}
\subsubsection{The structure of \idxname builds upon the $Mod.+C$}

\begin{figure}
\vspace{-1em}
  \centering
  \includegraphics[width=\linewidth]{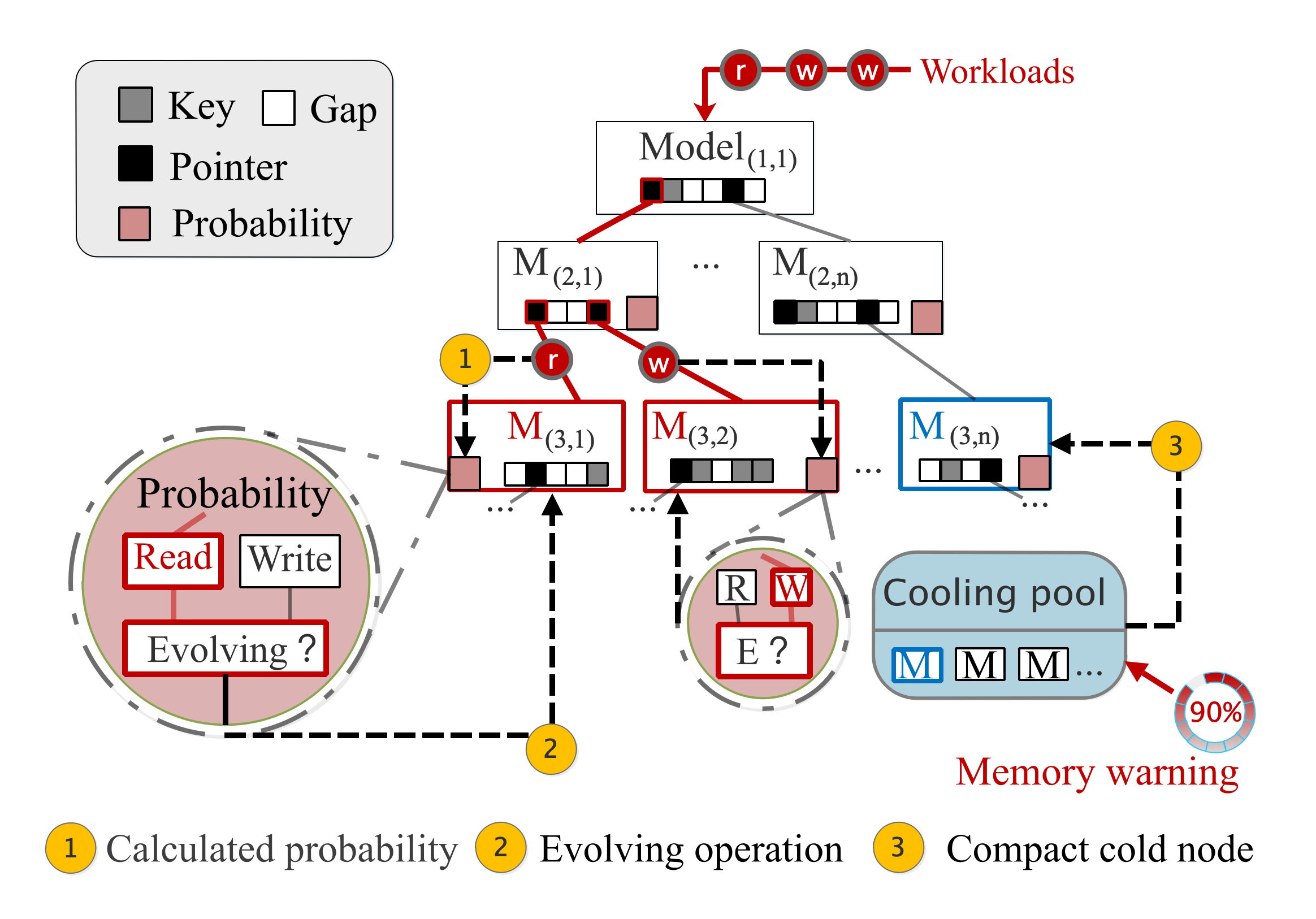}
  \vspace{-3em}
  \caption{The structure of \idxname builds upon the $Mod.+C$.}
  \vspace{-1.5em}
  \label{fig:figure 3}
\end{figure}

Based on our observation in Sections \ref{A concurrency-friendly learned index structure}, we have determined that the structure $Mod.+C$, i.e., the structure of LIPP, exhibits the highest scalability among the options considered. Consequently, we have opted to implement and evaluate our novel strategies utilizing $Mod.+C$ as the underlying index structure.
Note that in the subsequent context, \idxname refers to the structure built upon $Mod.+C$, as illustrated in \autoref{fig:figure 3}. Section \ref{Operations of SALI} and \ref{Coordination between different operations.} will introduce the operations of SALI and the coordination between different operations.

\vspace{-0.5em}
\subsubsection{Operations of \idxname} \ \
\label{Operations of SALI}

\begin{figure*}
\vspace{-2.5em}
\hspace{-1.5em}
  \centering
  \includegraphics[width=1.02\linewidth]{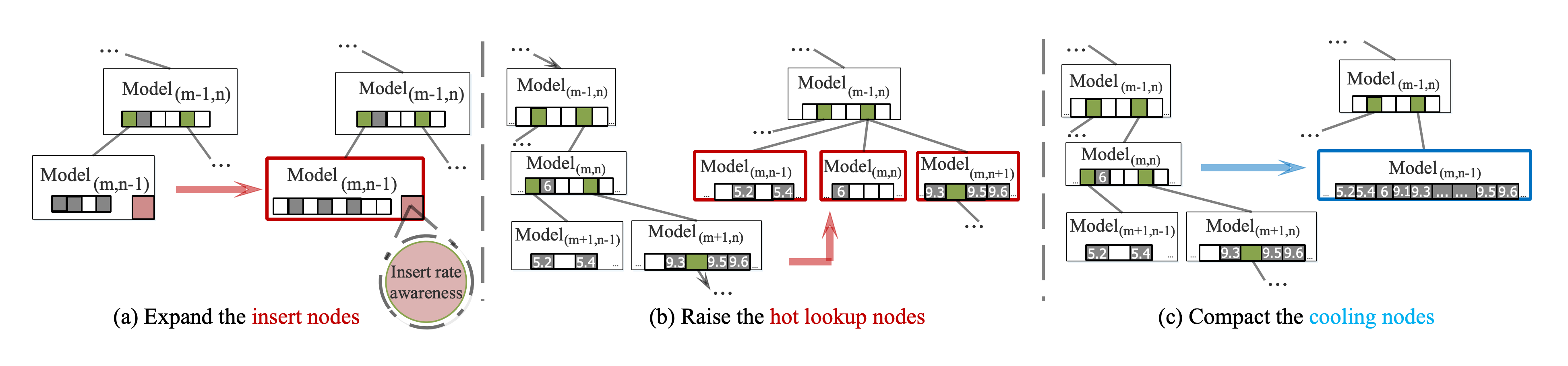}
  \vspace{-3.5em}
  \caption{The evolving strategies.}
  \vspace{-1em}
  \label{fig:figure 5}
\end{figure*}

1) Lookup operation: 
\idxname employs a linear model to accurately predict the position of the lookup key, except for cold nodes with errors.
The search is considered successful during a query if the key contained in the prediction slot is equal to the target key (Algorithm \ref{alg-readevolving}, line 4-6). 
Otherwise, it does not exist (Algorithm \ref{alg-readevolving}, line 7-9).
However, if the predicted slot is a pointer, the search continues in the node pointed to by the pointer (Algorithm \ref{alg-readevolving}, line 10-22). 
At this stage, the type of node needs to be determined.
For hot lookup nodes, SIMD~\cite{kim2010fast} is utilized to locate the node containing the target key, followed by a linear model search in this node (see \autoref{fig:figure 5}(b) and Algorithm \ref{alg-readevolving}, line 12-15). 
In contrast, for cold nodes, where the linear model prediction has an error, a binary search method covers the ``last mile'' (see \autoref{fig:figure 5}(c) and Algorithm \ref{alg-readevolving}, line 16-18). 
The rest of the nodes are searched directly using the linear model (Algorithm \ref{alg-readevolving}, line 19-20).

2) Insert operation:
Initially, the read operation algorithm is used to identify the appropriate location for inserting the key. 
If the key already exists at this position, a new storage space is created below, and the key is inserted in this space to handle conflicts (Algorithm \ref{alg-writeevolving}, lines 5-9). 
On the other hand, if the insertion position is a gap, the key is inserted directly (Algorithm \ref{alg-writeevolving}, lines 2-4).

% \vspace{-2.5em}

3) Evolving operation:
See Section~\ref{Evolving strategies} for details.

4) Building operation:
SALI adopts the structure of LIPP and therefore utilizes the construction algorithm of LIPP, i.e., fastest minimum conflict degree (FMCD)~\cite{wu2021updatable}.
In the \idxname and the realm of the learned index, other linear and even non-linear approximation algorithms are crucial and intriguing avenues for future research.

\vspace{-0.5em}
\subsubsection{Coordination between different operations.}
\label{Coordination between different operations.}

%lookup
Reads can proceed without acquiring the lock as long as \idxname verifies that the item being read (i.e., data or child pointer) has not been modified.

%insert
\idxname employs an optimistic locking mechanism for the target slot during concurrent writing.
Since the \idxname's structure ensures that only fine-grained locks are necessary to guarantee mutually exclusive writes, write conflicts are rare under a uniform workload.

%evolving
To prevent uncontrollable tail latency that may arise from prolonged evolution, we restrict the evolution process to nodes with less than one million keys. Our observations indicate that indexes with higher write rates require periodic rebuilding to maintain good performance.
Consequently, during periods of relative inactivity in the storage system, the entire index structure can be rebuilt, resulting in a flatter \idxname structure and improved performance.

Furthermore, when the node is evolving, we use the Read-Copy-Update (RCU) mechanism~\cite{mckenney2001rcu, siakavaras2020efficient} to prevent the blocking of read operations, i.e., reading the old version of data.
Following the evolution operation, \idxname utilizes the RCU to ensure all threads can access the new model. 
RCU barrier is a synchronization mechanism designed for concurrent systems, which enables all readers to access the new space in shared memory after evolving operation. 
In addition, to ensure that child nodes being read are not deleted during the evolution process, \idxname utilizes the epoch-based reclamation~\cite{fraser2004practical} that guarantees the safety of node pointers in a concurrent scenario.

% \vspace{-1em}
\begin{algorithm}[t]
        \small
	%\caption{Calculate $y = x^n$} 	
	\caption{\textcolor{black}{\idxname Lookup Operation}} 
	\label{alg-readevolving}
	\algorithmicrequire ~Target key: $k$.\ \ \ \ \ \ \ \ \ \ \ \ \ \ \ \ \ \ \ \ \ \ \ \ \ \ \ \ \ \ \ \ \ \ \ \ \ \ \ \ \ \ \ \ \ \ \ \ \ \ \ \ \ \ \ \ \ \ \ \ \ \ \ \ \ \ \ \ \ \ \ \ \ \ \ \ \ \ \ \ \ \\
	%暂时没有需要Output部分代码
	% \algorithmicensure ~The position of the key: $predicted\_slot$.\ \ \ \ \ \ \ \ \ \ \ \ \ \ \ \ \ \ \ \ \ \ \ \ \ \ \ \ \ \ \ \ \ \ \
	%改写成function函数形式
	\begin{algorithmic}[1]

	\Function{$lookup\_operation$}{$k$}
        \State $predicted\_slot \gets root.linear(k)$
        \While {(TRUE)}
            \If{$predicted\_slot.type == key$}
                \If{$predicted\_slot.data == k$}
                    \State \textbf{return} ($predicted\_slot.data$)
                \Else
                    \State \textbf{return} ($not\ \ found$)
                \EndIf
            \Else   \Comment{$predicted\_slot.type == pointer$}
                \State $nodes\_meta \gets predicted\_slot.data$
                \If{($nodes\_meta.type == hot\_lookup$)}
                    \State $nodes\_meta.linear \gets nodes\_meta.SIMD(k)$
                    \State $predicted\_slot \gets nodes\_meta.linear(k)$
                \EndIf
                \If{($nodes\_meta.type == cooling$)}
                    \State $pred\_slot\_pre \gets nodes\_meta.linear(k)$
                    \State $predicted\_slot \gets bi\_search (pred\_slot\_pre, k)$
                \Else   \Comment{$nodes\_meta.type == nomal$}
                    \State $predicted\_slot \gets nodes\_meta.linear(k)$
                \EndIf
            \EndIf
        \EndWhile
	\EndFunction
        \end{algorithmic}
\end{algorithm}

% \vspace{-2.5em}
\subsection{Evolving Strategies}
\label{Evolving strategies}

A more comprehensive adaptation strategy than simply retraining is required to adapt the learned index structure under various workloads. 
This part presents the design of an evolving strategy that focuses on three aspects to enhance the concurrency performance of the learned index. 
Note that this part only covers the evolving strategy, while the conditions and timing for triggering the evolving process will be discussed in Section~\ref{Probability Model}.
Next, we will briefly introduce the difference between evolving and retraining:

a) Retraining is a passive adjustment strategy used in updatable learned indexes to maintain their performance. Its main features are: 
1) it is driven by the deterioration of the index structure and cannot sense different workloads; 
2) it is triggered only by insert operations optimized exclusively for improving insert performance; 
and 3) it does not change the index structure in essence. 

b) Evolving, proposed in this paper, is a novel concept in learned indexes that includes the retraining function and improves the index's adjustment mechanism in different dimensions. Its main features are: 
1) it is an active adjustment strategy that perceives and is driven by the workload to improve index performance further; 
2) it can be triggered by any operation (e.g., read); 
and 3) it can ``evolve'' into a new structure type that is suitable for the current workload for both improving the read and insert performance. 
% Additionally, workload-aware adjustment makes learned index adjustment more cost-effective. For example, if a node is identified as cooling, its subtree structure will not be adjusted to improve throughput, even if it deteriorates significantly (i.e., becomes very deep). 

% The following part will introduce the three cases of \idxname's evolution.

\begin{algorithm}[t]
        \small
	%\caption{Calculate $y = x^n$} 	
	\caption{\textcolor{black}{\idxname Insertion Operation}} 
	\label{alg-writeevolving}
	\algorithmicrequire ~Target key: $k$.\ \ \ \ \ \ \ \ \ \ \ \ \ \ \ \ \ \ \ \ \ \ \ \ \ \ \ \ \ \ \ \ \ \ \ \ \ \ \ \ \ \ \ \ \ \ \ \ \ \ \ \ \ \ \ \ \ \ \ \ \ \ \ \ \ \ \ \ \ \ \ \ \ \ \ \ \ \ \ \ \ \\
	%暂时没有需要Output部分代码
	% \algorithmicensure ~The position of the key: $predicted\_slot$.\ \ \ \ \ \ \ \ \ \ \ \ \ \ \ \ \ \ \ \ \ \ \ \ \ \ \ \ \ \ \ \ \ \ \
	%改写成function函数形式
        
	\begin{algorithmic}[1]
        \Function{$insertion\_operation$}{$k$}
        \State $predicted\_slot \gets lookup\_operation(k) $
        \If{($predicted\_slot=NULL$)}
            \State insert($predicted\_slot, k$)
        \Else  \Comment{$predicted\_slot.type == key$}
                \State $old\_k \gets predicted\_slot.data$
                \State $pointer \gets predicted\_slot.type$
                \State $new\_node \gets predicted\_slot.data$
                \State insert($new\_node, old\_k, k$)
        \EndIf
        \EndFunction
        \end{algorithmic}
\end{algorithm}

% \vspace{-0.5em}
\subsubsection{The insert triggers the evolving operation.}

Most retraining methods, including \idxname, involve expanding the target node or its subtree. 
This expansion creates more gaps that can be used for inserting keys, thereby improving the overall insertion performance.
As shown in \autoref{fig:figure 5}(a), the gap array within the data node is increased from two to five to accommodate more keys. 
To achieve this, the FMCD algorithm~\cite{wu2021updatable} is used to expand the node by inputting all of the keys in the node and the desired space size after expansion (see Algorithm~\ref{alg-insert-evolving}, lines 1-4).

However, a fixed retraining expansion factor may not be sufficient to handle sudden increases in local insertions under a skewed workload, which can lead to a high number of insert conflicts in concurrent scenarios. 
To address this, the expansion factor should be \textbf{adaptively} adjusted based on the insertion rate to determine the optimal expansion size. 
Specifically, more gaps are reserved to enhance the insertion performance when the insertion rate increases. For more information, please refer to Equation (\ref{equation 1}) below:

\vspace{-1em}
\begin{equation}
\label{equation 1}
\begin{aligned}
n.expand\_size= \left \{
\begin{array}{ll}
 \gamma \times \frac{n.speed_t}{n.speed_{t-1}}\times n.build\_num,  &    \frac{n.speed_t}{n.speed_{t-1}} \geq 1 \\
 \gamma \times n.build\_num,                                        &\frac{n.speed_t}{n.speed_{t-1}} \textless 1
\end{array}
\right.
\end{aligned}
\end{equation}

Among them, $n$ refers to a specific node, $n.build\_num$ represents the size of the current node.  
$n.speed_t$ represents the accumulation rate at time $t$, indicating the insertion rate of new keys into a node at that specific time. This rate is determined by a probabilistic model, as described in Section~\ref{Probability Model}.
In Equation (\ref{equation 1}), a higher speed leads to a more significant expansion rate in the current operation, meaning that more gaps will be reserved compared to the previous expansion operation. 
The expansion factor $\gamma$ is defined as follows:

\begin{equation}
\label{equation 2}
\begin{aligned}
\gamma = \left \{
\begin{array}{ll}
 \theta, & n.build\_num \geq 1M \\
 2\theta, & n.build\_num \geq 100K\\
 5\theta, & n.build\_num \textless 100K
\end{array}
\right.
\end{aligned}
\end{equation}

Equation (\ref{equation 2}) reveals that nodes of varying sizes should have different expansion factors. 
Equation (\ref{equation 1}) demonstrates that smaller nodes ($n.build\_num$), need a more significant expansion factor to achieve adequate expansion. 
E.g., if the number of slots in two nodes is 4 and 8, respectively, both nodes would need to expand by 32 units, requiring expansion factors of 8 and 4, respectively.
The expansion base factor, denoted by $\theta$, can be dynamically adjusted based on varying workloads. 
For our evaluation, we set $\theta$ to its default value of 1.

\begin{figure*}
\vspace{-2em}
  \centering
  \includegraphics[width=\linewidth]{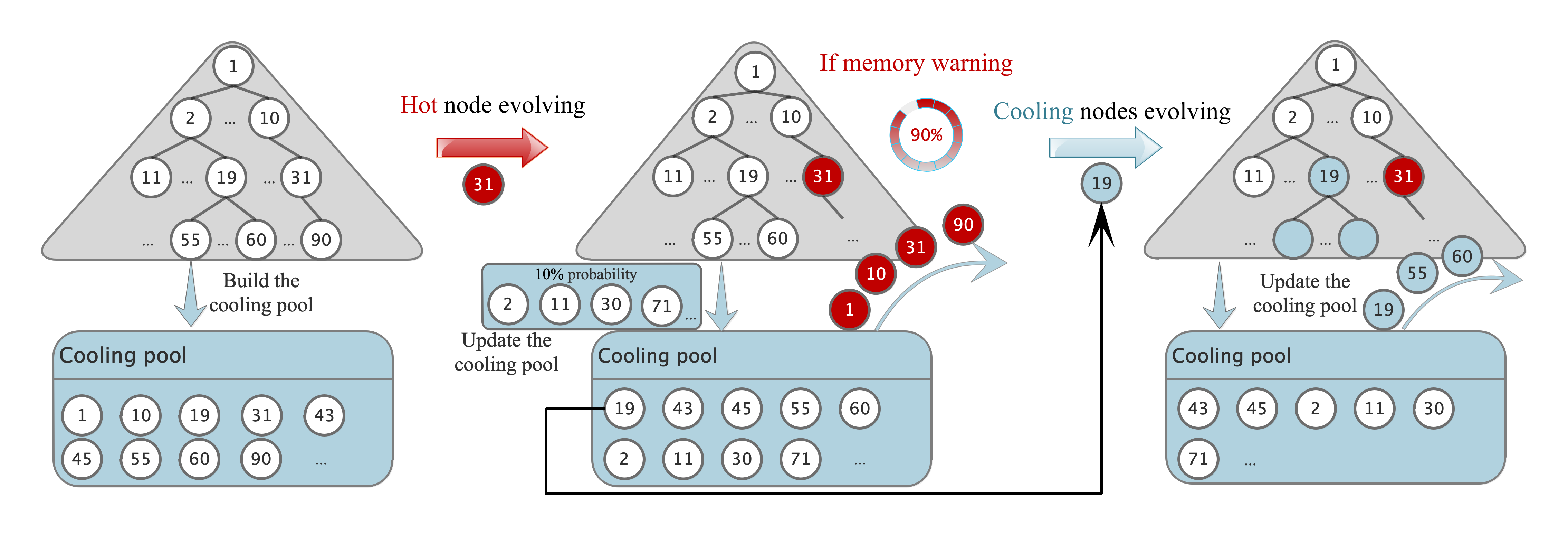}
  \vspace{-3.2em}
  \caption{The framework for identifying cold nodes.}
  \vspace{-1.3em}
  \label{fig:figure 4}
\end{figure*}

\begin{algorithm}[t]
        \small
	%\caption{Calculate $y = x^n$} 	
	\caption{\textcolor{black}{\idxname Insertion Evolving}} 
	\label{alg-insert-evolving}
	
	\algorithmicrequire ~Sequence $keys \{K_1,\cdots,K_n \}$ in a $Node$ and its $Subtree$. \ \ \ \ \ \ \ \ \ \ \ \\
	%暂时没有需要Output部分代码
	%\algorithmicensure $hello work$
	%改写成function函数形式
	\begin{algorithmic}[1]
        \Function{$Insert\_Node\_Evolving$}{$keys$}
        \State $segs.model \gets FMCD(keys, n.expand\_size)$
        \State $father\_slot$ \textbf{link} $segs.model $
        \EndFunction
	\end{algorithmic}
%	%\vspace{-0.3cm}
\end{algorithm}

\begin{algorithm}[t]
        \small
	%\caption{Calculate $y = x^n$} 	
	\caption{\textcolor{black}{\idxname Lookup Evolving}} 
	\label{alg-lookup-evolving}
	
	\algorithmicrequire ~Sequence $keys \{K_1,\cdots,K_n \}$ in a $Node$ and its $Subtree$. \ \ \ \ \ \ \ \ \ \ \ \\
	%暂时没有需要Output部分代码
	%\algorithmicensure $hello work$
	%改写成function函数形式
	\begin{algorithmic}[1]
        
	\Function{$Hot\_Lookup\_Node\_Evolving$}{$keys$}
        \State $gap\_array \gets calculate\_gap(keys) $
        \State $gap\_array \gets Top-k(gap\_array) $
        \For {$(gap\_array.size>0)$}
            \State $segment.key \gets split(gap\_array) $
            \State $segment.linear \gets approximation\_alg(segment.key) $
            \State $segs \gets segs.append(segment.key, segment.linear) $
        \EndFor
        \State $segs.line.slope \gets segs.line.slope \times n.expand\_size$
        \State $segs.model \gets insert(segment.key, segs.linear) $
        \State $father\_slot$ \textbf{link} $segs.model $
        \EndFunction
		
	\end{algorithmic}
%	%\vspace{-0.3cm}
\end{algorithm}

\begin{algorithm}[t]
        \small
	%\caption{Calculate $y = x^n$} 	
	\caption{\textcolor{black}{\idxname Cold Node Evolving}} 
	\label{alg-cooling-evolving}
	
	\algorithmicrequire ~Sequence $keys \{K_1,\cdots,K_n \}$ in a $Node$ and its $Subtree$. \ \ \ \ \ \ \ \ \ \ \ \\
	%暂时没有需要Output部分代码
	%\algorithmicensure $hello work$
	%改写成function函数形式
	\begin{algorithmic}[1]
  
        \Function{$Cooling\_Node\_Evolving$}{$keys$}
        \State $ segs.model \gets PLA\_algorithm(keys)$
        \State $father\_slot$ \textbf{link} $segs.model $
        \EndFunction
		
	\end{algorithmic}
%	%\vspace{-0.3cm}
\end{algorithm}

\subsubsection{The lookup triggers the evolving operation.}

We have developed an evolving strategy for hot read nodes to enhance concurrent read performance under skewed workloads further. 
Note that if the workload is uniform, \idxname can either choose to disable this evolving function or treat every node as a hot read node.
As depicted in \autoref{fig:figure 5}(b), we have designed a flat structure for hot reads nodes and their subtrees.
This structure flattens the nodes and promotes their levels as much as possible. 
Unlike the initial state where a single linear segment is linked under one slot, multiple segments can be linked under one slot after evolving. 
This flattening strategy reduces the tree height of the local hot structure.
Furthermore, \idxname can use SIMD instructions during lookup to quickly find which node the target key belongs to in the same layer.

We have developed a method to reconstruct structure, which has the effect of flattening the node and its subtrees (Algorithm~\ref{alg-lookup-evolving}). 
First, we sort all the nodes' keys, calculate the gap between two adjacent keys, and select the $top-k$ gap (Algorithm~\ref{alg-lookup-evolving}, line 2). 
Then, we split into $k-1$ segments based on these gaps and generate a linear model using the least squares algorithm (Algorithm~\ref{alg-lookup-evolving}, lines 3-8). 
According to Equation (\ref{equation 1}), we expand the slope of the linear model by a corresponding multiple to expand the corresponding space (Algorithm~\ref{alg-lookup-evolving}, line 9). 
Reserving the gap enables the CDF of the stored data to fit more easily on a line and improve lookup performance. 
Finally, we calculate the positions of all keys and insert them using the linear model after the slope expands. 
If there are still conflicts, we handle them similarly to \idxname's insertion conflict (Algorithm~\ref{alg-lookup-evolving}, line 10).

% Note that the linear model can be generated using any approximation algorithm, such as FMCD, depending on the CDF. 
% Choosing a better fitting method for different CDFs will be the focus of our future work. 
% Additionally, replacing the linear model with a nonlinear one may perform better and be the most critical research point for the learned index in the future.

% \vspace{-0.5em}
\subsubsection{Identify the cold node and trigger evolving operation.}

We developed a cold node-compression evolving strategy to optimize space usage in \idxname under skewed workloads. 
In addition to initially creating the \idxname index structure, we added a cooling pool space as illustrated in \autoref{fig:figure 4}.
During \idxname construction or each evolving operation, each node in the index has a 10\% probability of being chosen for inclusion in the cooling pool. 
When a node undergoes an evolving operation, that node, its subtrees, and all nodes above it are removed from the cooling pool.
At this stage, the nodes that remain in the cooling pool are considered temporarily cold.
We took inspiration for cold node design from~\cite{leis2018leanstore}.

Once each evolving operation finishes, \idxname checks whether the user-acceptable index size upper limit has been exceeded. 
If it has, \idxname selects the earliest-added node in the cooling pool for the compress operation and deletes it from the cooling pool until the space is reduced to meet the user-acceptable index size.

For cold nodes, we implemented a space compression strategy. 
As depicted in \autoref{fig:figure 5}(c), we cancel the reserved gaps to save space for cold nodes and their subtrees.
We use the PLA algorithm in PGM~\cite{ferragina2020pgm} to linearly approximate all keys in a cold node and generate the corresponding segment (Algorithm~\ref{alg-cooling-evolving}).

% \vspace{-0.5em}
\subsection{Probability Model}
\label{Probability Model}
% \vspace{-0.2em}

In order to ensure optimum performance, it is imperative that learned indexes monitor degradation statistics to initiate adjusting when necessary. 
Unfortunately, existing high-contention statistics techniques severely limit the scalability of learned indexes. Moreover, the implementation of the complete adjustment strategy, i.e., evolving presented in Section~\ref{Evolving strategies}, demands additional statistics, resulting in intolerable overhead in a concurrent scenario. 

To address this issue, we propose a probability-based strategy that employs a lightweight approach to maintain various statistics in \idxname to trigger evolving operations at a minimal cost.

Note that the fundamental concept behind the probability models is to leverage probabilities in simulating the accumulation of information.
For example, when simulating the cumulative number of inserted keys within a specified timeframe, we design a probability model based on the insertion rate and insertion time.
Furthermore, the geometric distribution can be utilized to simulate the accumulation of information such as insertion conflicts.

% \vspace{-0.5em}
\subsubsection{Probability model for triggering insert evolution.}
Most retrains are triggered by the deterioration of learned indexes caused by insert operations.
However, from an overall perspective of index performance, adjustments should be considered based on whether the local structure, following its adjustment, will continue to see the high-frequency insertion of new keys.  Such consideration can make the adjustment operation more advantageous, which achieves high amortized performance benefits.

Therefore, two conditions need to be considered to trigger insertion evolution:
1) The assessment of the frequency of new key insertions in a node and its subtree is crucial in determining whether an adequate number of keys are being inserted. 
The node's performance gains are higher after evolving if the frequency of new key insertions is high.
2) The escalation of conflicts within a node alongside the gradual increase in the number of newly inserted keys represents a critical aspect as it can be used to indicate the deterioration of the index.
Identifying deteriorating nodes is crucial, as only evolving such nodes will significantly improve performance.

Note that when a node satisfies only condition 1) and not condition 2), evolving is unnecessary because the insertion performance remains satisfactory.
When a node satisfies only condition 2) and not condition 1), the amortized performance benefit is low, and the evolving operation entails overhead costs. 
Therefore, satisfying both conditions simultaneously is a prerequisite for triggering the insert evolving operation. 
In the subsequent section, we provide a comprehensive analysis of the above two doctrinal conditions.

\textbf{Condition (1): the node accommodates a sufficient number of newly inserted keys.} 
To determine if this condition is met, we need to satisfy the following equation:

\begin{equation}
\label{equation 3}
  \frac{n.current\_num}{n.build\_num} \geq \beta
\end{equation}

$n.current\_num$ refers to the number of keys contained in the node at the end of the current insertion operation. 
$n.build\_num$ indicates the number of keys in the node when the last ``evolving'' operation was performed. 
The tolerance coefficient $\beta$ specifies the maximum amount of data that can be inserted into the node before it needs to be adjusted.
As a general guideline, we set $\beta=2$.

Since each insertion thread must maintain the cumulative variable $current\_num$, conflicts may arise. 
To address this issue, we propose a lightweight probability model.

First, we multiply the insertion rate by the timestamp difference to get the total amount of inserted new keys during this period and put it into Equation (\ref{equation 3}) to get:

\begin{equation}
\label{equation 4}
  \frac{[n.speed_t\times(n.current\_time-n.build\_time)]+n.build\_num}{n.build\_num} \geq \beta
\end{equation}

The variable $n.build\_time$ represents the timestamp corresponding to state $n.build\_num$, while $n.current\_time$ represents the current timestamp. 
The estimated insertion rate, denoted as $n.speed_t$\footnote{We assign a specific value to $speed_1$.}, is calculated using Equation (\ref{equation 5}), which takes the quotient of the total number of insertions and the difference in timestamp from the previous period. 
Therefore, we can estimate the value of $n.current\_num$ using $n.speed_t\times(n.current\_time-n.build\_time)]+n.build\_num$ at time $t$.

\begin{equation}
\label{equation 5}
  \frac{n.current\_num - n.build\_num}{n.current\_time-n.build\_time} = n.speed_{t+1}
\end{equation}

Additionally, we define the cumulative probability within a node as $P_{acc}$ (see Equation (\ref{equation 6})), as obtained through the transformation of Equation (\ref{equation 4}). When $P_{acc}=1$, condition (1) is met. When $P_{acc}<1$, we determine whether the evolving adjustment is necessary based on a Bernoulli experiment. If the experiment is successful, condition (1) is met; otherwise, it is not met. 

\begin{equation}
\label{equation 6}
  P_{acc} = \frac{[n.speed_t\times(n.current\_time-n.build\_time)]}{(\beta-1) \times n.build\_num}
\end{equation}

Note that the calculation of Equation (\ref{equation 5}) may result in $n.speed$ being zero. 
In such cases, the cumulative probability computed by Equation (\ref{equation 6}) will always be zero, preventing any further changes in $n.speed$. 
To resolve this issue, we introduce a reconciling variable $\epsilon$ in the numerator of Equation (\ref{equation 6}).
Expressly, we set $\epsilon = path\_size/1000$, where $path\_size$ denotes the path length from the root node to the current node.
The final Equation of the cumulative probability model, Equation (\ref{equation 7}), determines whether the node accommodates a sufficient number of newly inserted keys.

\begin{equation}
\label{equation 7}
  P_{acc} = \frac{[n.speed_t\times(n.current\_time-n.build\_time)]+\epsilon}{(\beta-1) \times n.build\_num}
\end{equation}

\textbf{Condition (2): the node accommodates an adequate number of conflicts.}
We calculate the ratio of insertion conflicts to the total number of insertions:

\begin{equation}
\label{equation 8}
  \frac{n.conflict\_num}{n.current\_num-n.build\_num} \geq \alpha
\end{equation}

The variable $n.conflict\_num$ denotes the total number of conflicts in the node resulting from the insertions between the last evolving operation and the current state. 
Meanwhile, the conflict tolerance coefficient is denoted by $\alpha$, which we typically set to 0.1 as a rule of thumb.
Similar to Condition (1), in a concurrent scenario, we need to develop a probability model to estimate the number of conflicts to avoid blocking threads according to Equation (\ref{equation 8}).

For an evolving operation to occur, the node must have sufficient newly inserted keys. Therefore, we use $(\beta-1)  \times n.bulid\_num$ to estimate the number of new insertions according to Equation (\ref{equation 3}) when there are enough conflicts to cause evolution:

\begin{equation}
\label{equation 9}
  (n.current\_num-n.build\_num) \approx (\beta-1) \times n.bulid\_num
\end{equation}

By substituting Equation (\ref{equation 9}) into Equation (\ref{equation 8}), we obtain Equation (\ref{equation 10}):

\begin{equation}
\label{equation 10}
  n.conflict\_num \geq \alpha \times (\beta-1) \times n.bulid\_num
\end{equation}

%实验的第一张图
\begin{figure*}
\vspace{-1.2em}
% \hspace{-3em}
  \centering
  % \includesvg[width=1\linewidth]{figure_8.svg}
  \includegraphics[width=1\linewidth]{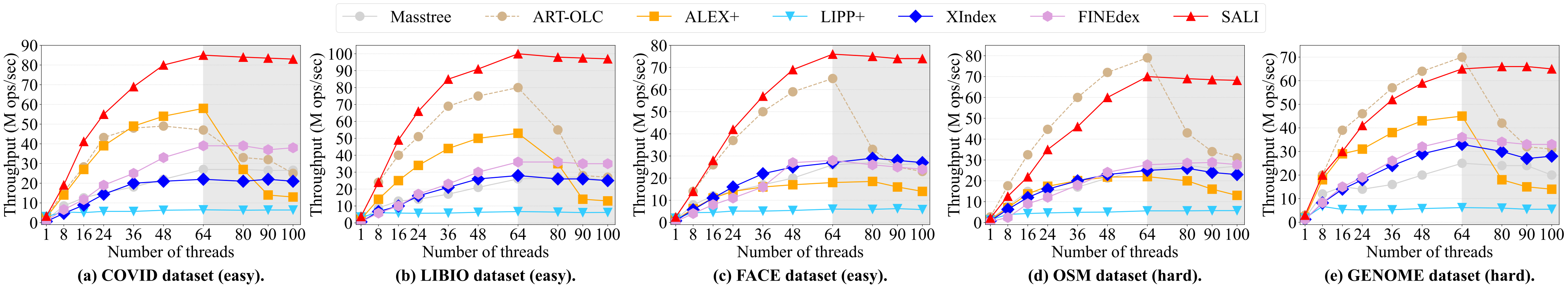}
  \vspace{-2.5em}
  \caption{The indexes scalability on write-only workloads. The grey area indicates that the threads number exceeds the maximum logical cores number. Extended plots with all evaluations are available here:~\cite{appendix}}
  \vspace{-1.6em}
  \label{fig:figure 8}
\end{figure*}

When an insertion causes a conflict, we set the conflict adjustment probability to $P_{conflict}$. 
Using the expectation of the geometric distribution, we can estimate that the expected number of conflicts that trigger evolving after the conflict is $\frac{1}{P_{conflict}}$, i.e., $\frac{1}{P_{conflict}} \approx n.conflict\_num$. 
Thus, whenever a conflict occurs in a node, we trigger the probability model specified in Equation (\ref{equation 11}) to determine if the model deteriorates and requires evolving. 

\begin{equation}
\label{equation 11}
  P_{conflict} = \frac{1}{\alpha \times (\beta-1)  \times n.bulid\_num}
\end{equation}

\textbf{Application in \idxname:} 
In \idxname, we only compute probabilities when a conflict occurs to minimize overhead. 
We determine whether $P_{conflict}$ is triggered; if so, we proceed to determine whether $P_{acc}$ is also triggered.  
If both conditions are met, the evolving operation is necessary to adjust the insertion structure of \idxname.
% Note that, to prevent frequent evolving operations, only when the number of keys in a node is greater than 64 will there be a probability model. 
% This rule also applies to Section 3.2.2 and Section 3.2.3

\subsubsection{Probability model for triggering lookup evolution.}

We define the probability that a target node is identified as a \textbf{h}ot read node due to a \textbf{l}ookup operation, denoted as $P_{hl}$, that is a hyperparameter that can be set to an appropriate value.
Whenever the lookup operation encounters a node, we can check whether $P_{hl}$ is triggered for that node. 
If $P_{hl}$ is triggered, we consider the node and its subtree as a hot lookup structure.

In addition to the probability $P_{hl}$, the following conditions for setting the read trigger probability also need to be considered:

(1) The evolving operation has not been triggered by lookup operations on the node for a prolonged period of time.

% •	The evolving operation has not triggered the node by the insert operation for a long time.

(2) The rate at which the node accumulates data ($n.speed_t$) through insertions is not slow.

For condition (1), if the last evolve operation of a node was triggered by a hot lookup, it means that no insert operation has triggered the node to evolve since then, 
i.e., the node has not severely deteriorated, and the number of new 
insertion keys are likely to be few.
In this case, we can adjust $P_{hl}$ to a smaller value to prevent frequent evolving, i.e., $P_{hl} = P_{hl}\times \lambda$, where $\lambda$ is a penalty coefficient.

For condition (2), in addition to $P_{hl}$, we introduce the probability $P_{acc}$, as defined in Equation (\ref{equation 7}). 
If a large number of new keys are inserted since the last evolving operation, it suggests that a new round of evolving operations may be necessary.

% To sum up, to determine whether it is a hot read node, it is necessary to ensure that $P_{hl}$ is triggered and $P_{acc}$ is triggered simultaneously.

\textbf{Application in \idxname:} 
We generate a $skip\_counter$ in each lookup thread-local, which maintains the number of lookup operations.
Upon execution of a lookup operation, the $skip\_counter$ is incremented by 1. 
After every 10 lookup operations, a Bernoulli experiment is conducted to determine whether $P_{hl}$ is triggered.
If $P_{hl}$ is triggered, the system verifies whether $P_{acc}$ is also triggered. 
If $P_{acc}$ is triggered, \idxname proceeds with the evolving operation.

% \vspace{-0.5em}
\section{Evaluation}
\label{evalution}

This section conducts a comprehensive evaluation of \idxname. 
Section~\ref{Experimental Setup} describes the experimental setup. 
Section~\ref{Overall Results} compares \idxname's performance with that of several state-of-the-art concurrent learned indexes and traditional indexes using various datasets and thread counts. 
Section~\ref{Evolving evaluation} evaluates \idxname under skewed workloads. 
Finally, Section~\ref{Ablation study} conducts an ablation study on \idxname. 

% \vspace{-0.5em}
\subsection{Experimental Setup}
\label{Experimental Setup}

All experiments are conducted on a two-socket server with two 16-core Intel Xeon Gold 6242 @2.80GHz CPUs (hyper-threading to 64 threads) and 384GB of DRAM. We implemented \idxname with $\sim$4k LOC of C++.

%实验第二张图
% \vspace{-0.8em}
\begin{figure}
% \hspace{-3em}
  \centering
  \includegraphics[width=\linewidth]{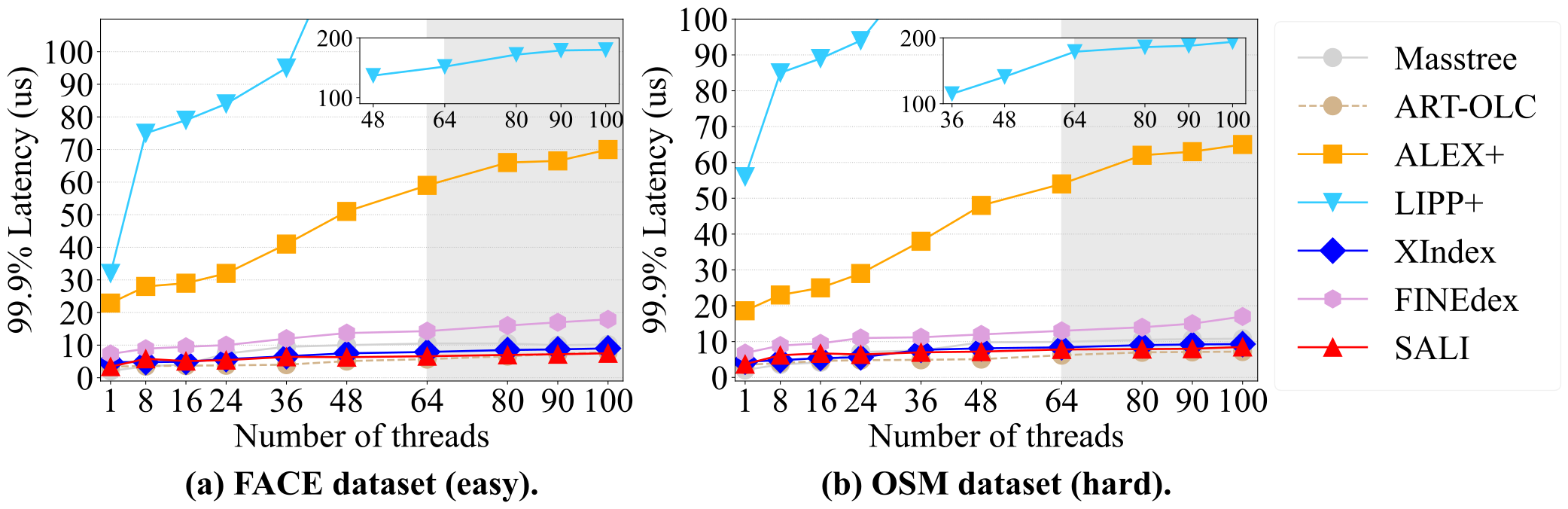}
  \vspace{-2.5em}
  \caption{The latency of indexes on write-only workloads.}
  \vspace{-2.2em}
  \label{fig:figure 8.1}
\end{figure}

%实验第三张图
\begin{figure*}
\vspace{-1.2em}
% \hspace{-3em}
  \centering
  % \includesvg[width=1\linewidth]{figure_9.svg}
  \includegraphics[width=\linewidth]{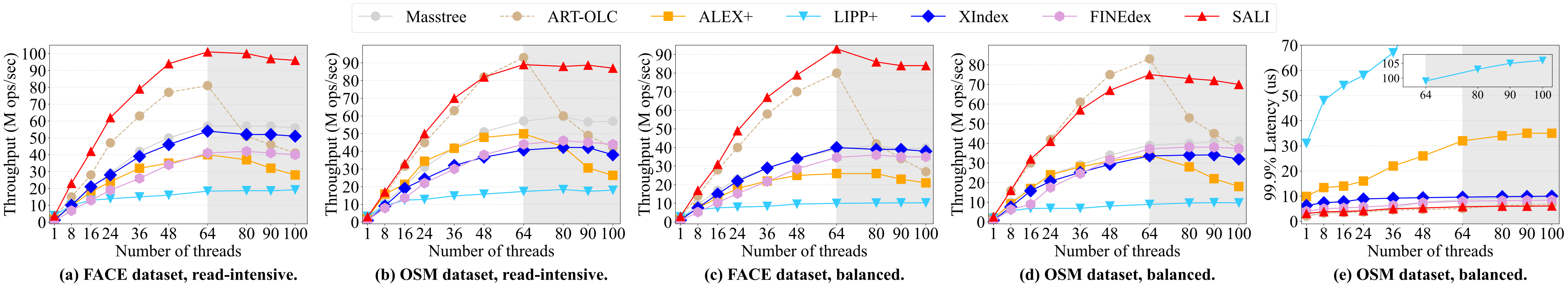}
  \vspace{-2.5em}
  \caption{The indexes scalability on read-write workloads. Extended plots with all evaluations are available here:~\cite{appendix}}
  \vspace{-1.3em}
  \label{fig:figure 9}
\end{figure*}

% \vspace{-0.5em}
\subsubsection{Baselines}
We benchmarked \idxname against six baselines. 
1) Masstree~\cite{masstree}, a hybrid index structure of B+Tree and Radix Tree; 
2) ART-OLC~\cite{2016artolc}, an exemplary concurrency implementation of the Adaptive Radix Tree (ART)~\cite{art2013adaptive}. 
3) ALEX+~\cite{gre}, an exemplary concurrency implementation of the ALEX~\cite{ding2020alex}. 
4) LIPP+~\cite{gre}, a concurrency implementation of the LIPP~\cite{wu2021updatable}. 
5) XIndex~\cite{tang2020xindex}, a first attempt to design a concurrent learned index.
6) FINEdex~\cite{li2021finedex}, a fine-grained updated concurrent learned index.

% \vspace{-0.5em}
\subsubsection{Datasets}
We selected several real datasets from SOSD~\cite{marcus2020benchmarking} and GRE~\cite{gre} benchmarks. 

•	COVID: Tweet ID with tag COVID-19~\cite{lopez2021covid} (Uniformly sampled).

•	FACE: Facebook user ID~\cite{facebook}.

•	LIBIO: Repository ID from libraries.io~\cite{gre}.

•	OSM: OpenStreetMap locations~\cite{marcus2020benchmarking} (Uniformly sampled).

•	GENOME: Pairs of locations on human chromosomes~\cite{rao20143genome}.

% •	Wise: Partition key from the WISE data~\cite{wright2010wise}.

Note that according to paper~\cite{gre}, the OSM and GENOME datasets are considered to be of ``hard'' difficulty for learned indexes, as fitting a Cumulative Distribution Function (CDF) on these datasets is challenging. 
Relatively, fitting remaining datasets with a CDF is comparably easier.

% \vspace{-0.5em}
\subsubsection{Workloads}
We design workloads to generate requests using the aforementioned datasets. 
To achieve this, we randomly shuffle all 200 million keys for each dataset and issue insert and lookup requests based on the following ratios:

•	Read-Only: Load all 200M keys and randomly search 800M.

•	Read-Intensive (20\% insert): Load 100M random keys and perform 80\% search $\&$ 20\% insert, i.e., insert all the remaining keys.

•	Balanced (50\% insert): Load 100M random keys and perform 50\% search $\&$ 50\% insert, i.e., insert all the remaining keys.

•	Write-Only: Load 100M keys and insert 100M keys.

•	Hot-read-A (100\% Read): Load 200M keys. Select 1/10 of these 200M (20M) as hot read keys and execute five rounds of read operations on these 20M keys, i.e., 100M read operations.

•	Hot-read-B (16\% insert): Perform an additional insert operation of 20M keys based on the keys from Hot-read-A.

•	Hot-write (100\% Insert): Randomly select one-eighth consecutive data in the 200M data as the hot insert, and insert it after loading the remaining data.

% Please note that the Hot-write workload will be performed eight times and averaged to obtain the final result, whereas the other workloads will be performed three times.

We repeated 10 and 5 experiments for Hot-write and other workloads, respectively, excluding the lowest and the highest measure, and reported the average of the results.
Between each measurement of experiments, we wiped caches and re-loaded the data to avoid intermediate results.

% \vspace{-0.8em}
\subsection{Overall Results}
\label{Overall Results}
% \vspace{-0.3em}

This section evaluates the \idxname's overall performance against the SOTA indexes that support concurrency. 
In this experiment, uniform workloads were used. To ensure fairness, \idxname turns off the judgment and evolution modules temporarily of hot read nodes but reserves the evolving operations triggered by the insertion on \idxname, as all indexes require maintaining statistical information and performing retraining operations when inserting data.

% \begin{figure*}[h]
% % \hspace{-3em}
%   \centering
%   \includesvg[width=1\linewidth]{figure 10.svg}
%   \vspace{-2.5em}
%   \caption{The indexes scalability on balanced workloads.}
%   \vspace{-1.5em}
%   \label{fig:figure 10}
% \end{figure*}

% \vspace{-0.5em}
\subsubsection{Write-only workloads}

\autoref{fig:figure 8} shows the concurrent performance evaluation of various learned indexes when executing the write-only workload on different datasets. 
The triggered retraining is directly executed in the foreground by the thread responsible for triggering. 
These threads are represented by the numbers specified on the x-axis.
All indexes except LIPP+ benefit from increasing threads, but performance drops when the number of threads exceeds the logical thread count.

In easy datasets, namely COVID, LIBIO, and FACE, \idxname performs better in terms of throughput overall. 
Compared to the learned index ALEX+ and traditional index ART-OLC, \idxname exhibits the best scalability. 
ALEX+ and ART-OLC exhibit a sharp performance drop when the number of threads exceeds 60, whereas \idxname maintains a high and stable performance. 
In the COVID dataset, \idxname outperforms the best two baselines, i.e., ALEX+ and ART-OLC, by up to 47\% and 73\% at the highest solution, respectively, and the advantage continues to expand beyond 60 threads.

In hard datasets, \idxname outperforms other learned indexes by up to a factor of 2.5x to 10x, 
% While it is challenging for the learned index to fit the CDF accurately, \idxname’s performance can still approach that of ART-OLC.
% However, ART-OLC's also performance drops sharply when more than 60 threads.
as \idxname can accurately fit complex CDF, so that its performance can match ART-OLC and still maintain performance even exceeding the number of logical threads, where, in contrast, the performance of ART-OLC sharply declined.

It is noteworthy that ALEX+ and ART-OLC exhibit significant performance degradation when threads exceed 60. 
This issue is due to the coarse-grained lock, resulting in severe thread blocking when insertion (as discussed in Section~\ref{A concurrency-friendly learned index structure}).
% e.g., the shift operation when ALEX+ inserts will cause the entire leaf node to be locked, blocking other threads that need to insert this leaf. This situation of ART-OLC also occurs. In addition, ALEX+ has a significant prediction error due to its difficulty fitting, especially on the hard dataset. Therefore ALEX+ needs a long "last mile" search, resulting in memory bandwidth exhaustion, and performance will suffer severely at low thread counts. 
% In fact, through experimental observation, we found that when the number of threads reaches 24, ALEX+'s memory bandwidth is exhausted~\cite{gre}.

Additionally, the buffer-based insert strategy used by XIndex and FINEdex results in a large lookup error, and frequent ``last mile'' searches result in poor scalability. 
LIPP+ employs a high-contend method for all nodes to statistic information, leading to severe blocking of insertion threads and cache-line ping-pong, which complete loss of scalability in a concurrent scenario.

\autoref{fig:figure 8.1} shows the 99.9\% tail latency on FACE and OSM datasets. 
As threads increase, ALEX+ and LIPP+ exhibit significant increases in tail latency. 
ALEX+ has coarse-grained locks, which cause thread blocking during insertion and retraining operations, resulting in a significant increase in tail latency. 
LIPP+ requires joint maintenance of statistical information by different threads, leading to an increase in thread blocking and a significant increase in tail latency. 
In contrast, the other indexes do not show noticeable increases, and \idxname maintains the lowest in most settings.

\textbf{Insight 1: Concurrent insertion often faces three challenges: a) high-contend statistics maintenance causing thread blocking; b) coarse-grained write locks leading to thread blocking; c) write amplification and lookup errors causing memory bandwidth exhaustion. 
\idxname efficiently addresses them, leading to exceptional scalability, especially with hard datasets and a high number of threads.}

% \vspace{-0.5em}
\subsubsection{Read-write workloads}

The performance of different indexes under read-intensive and balanced workloads is shown in \autoref{fig:figure 9}.
% , where \idxname demonstrates excellent scalability.

\autoref{fig:figure 9}(a,b) exhibit that \idxname outperforms other indexes under read-intensive workloads, especially in the easy dataset. Compared to ART-OLC and ALEX+, \idxname improves the performance by up to 37\% and 55\%, respectively, under 60 threads.

\autoref{fig:figure 9}(c,d) illustrate the performance under the balanced workload. 
\idxname maintains high scalability. 
Other indexes' performance also exhibits similar to the read-intensive workloads.

Nevertheless, LIPP+ remains unscalable.
And when the number of threads exceeds 60, the performance of ALEX+ and ART-OLC degrades significantly.
The aforementioned indicates that even with a low proportion of write operations (20\% insert), the indexes experience a bottleneck under ultra-high threading conditions.

In \autoref{fig:figure 9}(e), the tail latency of the index under the balanced workload is illustrated. 
Similar to \autoref{fig:figure 8.1}, ALEX+ and LIPP+ exhibit higher tail latencies compared to the other indexes. 
However, as the read ratio increases, the tail latency of both indexes decreases. 
Notably, \idxname maintains consistently low tail latency throughout.

% \begin{figure}
% % \hspace{-3em}
% % \vspace{-1.3em}
%   \centering
%   \includesvg[width=\linewidth]{figure 15.svg}
%   \vspace{-2.5em}
%   \caption{bulk time and range query performance.}
%   \vspace{-2.1em}
%   \label{fig:figure 15}
% \end{figure}

\textbf{Insight 2: 
\idxname exhibits outstanding scalability under workloads that involve insertion operations, even under hard datasets. 
Conversely, other indexes face scalability bottlenecks, even with a low proportion of insert operations.}

\begin{figure*}
% \vspace{-1.1em}
% \hspace{-3em}
  \centering
  % \includesvg[width=1\linewidth]{figure_11.svg}
  \includegraphics[width=1\linewidth]{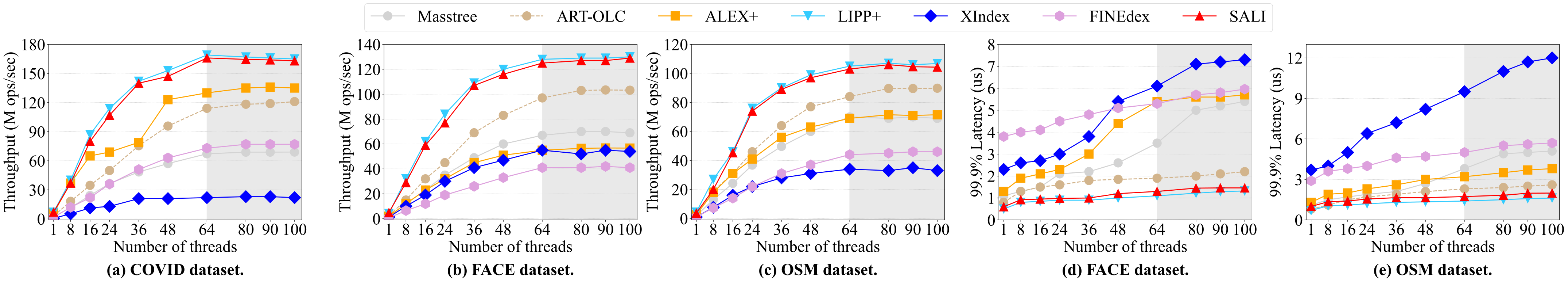}
  \vspace{-2.5em}
  \caption{The indexes scalability on read-only workloads. Extended plots with all evaluations are available here:~\cite{appendix}}
  \vspace{-1.8em}
  \label{fig:figure 11}
\end{figure*}

\begin{figure*}
% \hspace{-3em}
\vspace{-0.5em}
  \centering
  \includegraphics[width=\linewidth]{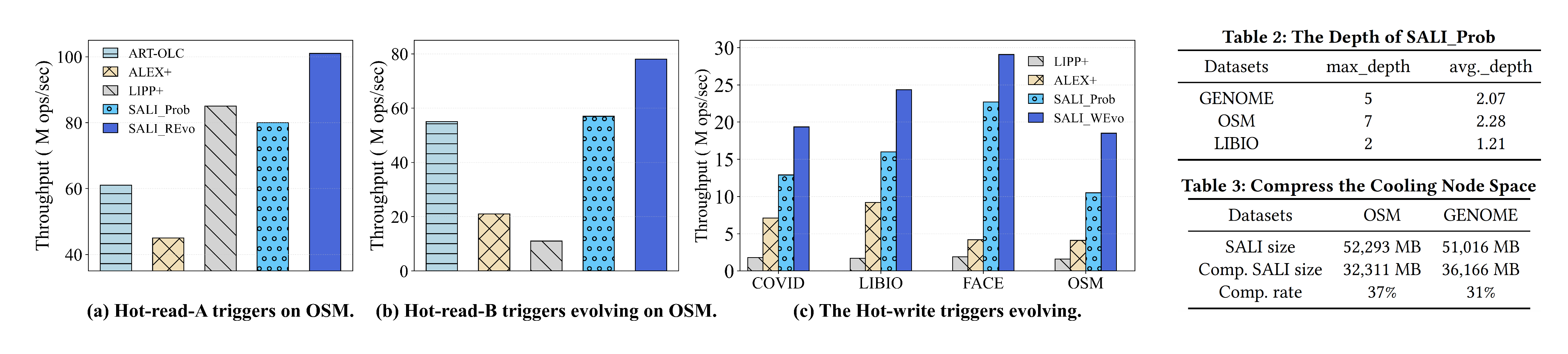}
  \vspace{-3.6em}
  \caption{The performance of the evolving strategy.}
  \vspace{-1.7em}
  \label{fig:figure 12}
\end{figure*}

% \vspace{-0.5em}
\subsubsection{Read-only workloads}

\autoref{fig:figure 11} presents the evaluation of the read-only workload. \idxname and LIPP+ outperform other indexes in both easy and hard datasets, as they adopt a model-based insert + chain structure, which enables accurate lookups.

LIPP+ does not require high-contend maintaining statistics in read-only scenarios, while \idxname needs to identify hot and cold nodes, which adds a slight overhead. 
Therefore, \idxname's performance is slightly lower than that of LIPP+.

ALEX+ exhibits poor lookup performance due to frequent ``last mile'' lookups.
XIndex and FINEdex perform unsatisfied in general due to serious lookup errors.
% ART-OLC loses competitiveness in easy and hard datasets.
ART-OLC does not have high superiority with the read-only workload due to its higher tree height in comparison to the learned index.

\autoref{fig:figure 11}(d,e) depict the tail latency on FACE and OSM datasets. 
XIndex and FINEdex perform worse due to the severe lookup errors. 
In contrast, \idxname and LIPP+ have lower tail latency than ALEX+, as they do not suffer from any lookup errors.

\textbf{Insight 3: 
All indexes benefit from hyperthreading under the read-only. 
Among them, \idxname and LIPP+ deliver the best performance due to accurate lookup capability, which is essential for improving query performance.
}

% \vspace{-0.5em}
% \subsubsection{Index building and range queries}

% \autoref{fig:figure 15}(a) presents the time required for bulk loading 100M keys by various learned indexes. 
% Notably, \idxname's build time is significantly lower than that of ALEX+ and XIndex. 
% This is because \idxname creates new nodes for conflicting keys and does not require key movement to maintain gaps, as in ALEX+. 
% \idxname needs to create a cooling pool, resulting in a slightly longer build time than LIPP+.

% \autoref{fig:figure 15}(b) presents the evaluation of a range query of 100 keys with 48 threads. 
% \idxname outperforms XIndex and FINEdex but falls short of ALEX+. 
% This is because \idxname's node layout, which resembles that of a B-tree, contains more gaps and interleaves child pointers and data in the node array. 
% Consequently, scan on the array encounters many branches. 
% To mitigate this issue, we make preliminary optimizations by compressing the hot scan node (similar to the cold nodes compression designed in \autoref{fig:figure 5}(c)) to remove the gap and store the data in a node, as illustrated in \idxname+Comp. in \autoref{fig:figure 15}(b), assuming that we know the hot scan information. 
% The results show that this approach outperforms other baselines. 
% Our future research will focus on designing more robust methods for identifying hot scan nodes, building on the prerequisites provided by the lightweight models.

% \vspace{-0.7em}
\subsection{Evolving Evaluation}
\label{Evolving evaluation}
% \vspace{-0.2em}

\subsubsection{Evolving triggered by hot read}
\label{Evolving triggered by hot read}

\autoref{fig:figure 12}(a,b) compare the performance of the learned index with 48 threads in two hot-read workloads on the OSM dataset. 
\idxname\_Prob refers to \idxname with only the probability model, while \idxname\_REvo uses an evolving strategy triggered by read.
The figure exhibits that \idxname\_REvo outperforms \idxname\_Prob by 27\% and 36\% in the two hot-read workloads, respectively.
In the Hot-read-A workload, \idxname\_REvo has surpassed the performance of LIPP+.

Moreover, \idxname\_REvo performs exceptionally well in OSM and GENOME due to the reduction in subtree height of hot read nodes, effectively increasing read performance.
Table 2 (in \autoref{fig:figure 12}) shows that the \idxname\_Prob's depth on the two hard datasets is up to 5 and 7, respectively. 
Therefore, the evolution of the hot node will flatten the node to optimize read performance. 
However, under the easy datasets, such as LIBIO, the average depth of \idxname\_Prob is only 1.2, so there exists a negligible improvement when using \idxname\_REvo as there is not enough depth to reduce.

Note that as SALI adopts the LIPP structure, the majority of keys are stored in the root node and upper levels, while the deeper subtree contains fewer keys. 
Therefore, during the read evolution, we found that connecting two nodes in one slot yields the best performance while connecting more nodes would increase the overhead of indexing fewer keys, rendering the evolution strategy ineffective.
In this case, we directly determine which of the two nodes the target key belongs to based on the maximum value of the nodes, which is more efficient than the SIMD approach. 
In Section~\ref{discussion}, we further discuss the applicability and limitations of read evolving.

\textbf{Insight 4: 
In skewed workloads, the hot-read evolving can significantly improve read performance when the subtree of the hot-read node is deep. 
Flattened tree structures under easy datasets do not require evolving.}

% \begin{table}[hb]% h asks to places the floating element [h]ere.
% \setlength{\abovecaptionskip}{0pt}%    
% \setlength{\belowcaptionskip}{10pt}%
%   \caption{The Depth of \idxname\_Prob}
%   \label{tab:table 2}
%   \begin{tabular}{cccc}
%     \toprule
%     Datasets & max\_depth & avg.\_depth\\
%     \midrule
%     GENOME    &   5  &   2.07   \\
%     OSM       &   7  &   2.28   \\
%     LIBIO     &   2  &   1.21   \\
   
%   \bottomrule
% \end{tabular}
% \end{table}

\vspace{-0.7em}
\subsubsection{Evolving triggered by the hot insert.}

\autoref{fig:figure 12}(c) presents the evaluation of Hot-write workloads with 48 threads. 
\idxname\_WEvo includes both the probability model and evolving strategy triggered by insertion.
\idxname\_Prob only consists of the probability model and an adjustment strategy equipped with a fixed $n.speed_t$ (as described in Equation (1)), similar to the adjustment method used by existing learned indexes, i.e., the expansion coefficient is fixed to expand the corresponding node during adjustment.

\idxname\_WEvo outperforms \idxname\_Prob by 32\% to 80\%, indicating that the evolving strategy significantly impacts hot write nodes. 
However, the performance of the index structure during the hot write workload is not as good as that of the uniform workload. 
This is because hot writes cause the local structure of the index to deteriorate continuously and require frequent retraining operations. 
Additionally, frequent local writes in ALEX+ and ART-OLC index structures with coarse-grained write locks increase thread blocking and adversely affect insertion performance.

Nonetheless, \idxname\_WEvo can adaptively evolve the hot node based on the insertion rate, i.e., $n.speed_t$ (see Section~\ref{Evolving strategies}). 
When the insertion rate becomes faster, more slots are reserved to ensure excellent insertion performance through the expand operation (see Equation (1)), which significantly reduces the number of retraining operations and improves overall performance.

\textbf{Insight 5: 
% Skewed workloads can cause frequent writes to the local index structure, leading to significant deterioration. 
% However, \idxname can effectively handle high local insertion rates by dynamically reserving more gaps to maintain a high throughput.
Skewed workloads with hot writes often lead to significant performance deterioration. \idxname can effectively analyze and process hot insertion by dynamically reserving more gaps to maintain a high throughput.
}

\begin{figure}
% \hspace{-3em}
\vspace{-0.5em}
  \centering
  % \includesvg[width=0.97\linewidth]{figure_13.svg}
  \includegraphics[width=0.97\linewidth]{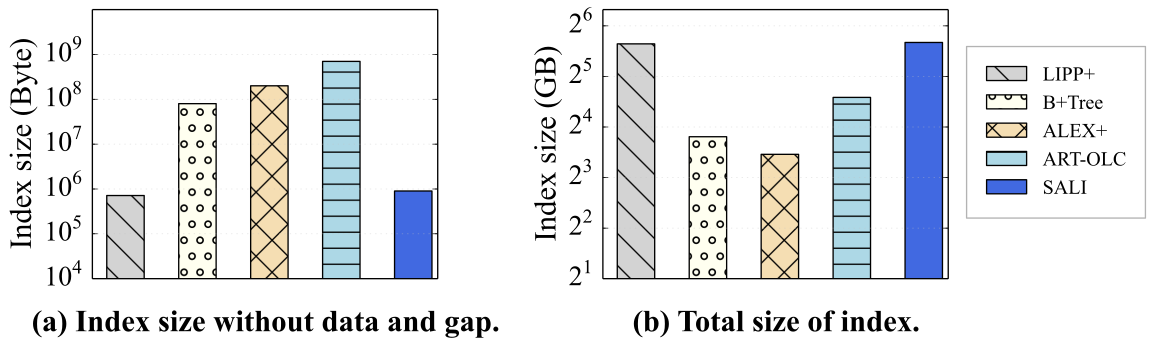}
  \vspace{-1.5em}
  \caption{The size of indexes.}
  \vspace{-2em}
  \label{fig:figure 13}
\end{figure}

% \vspace{-0.7em}
\subsubsection{Evolving triggered by cold nodes}

\autoref{fig:figure 13} illustrates the size of learned indexes on OSM. 
Notably, the size of learned indexes on OSM is about 1.5 times larger than that of the easy datasets. 
In particular, \autoref{fig:figure 13}(a) shows the internal structure size of the indexes, indicating that \idxname incurs the smallest space overhead. 

Note that the space overhead of the key-value pair could overshadow that of the index.
\autoref{fig:figure 13}(b) presents the sum of the index's internal structure, the gap, and the key-value pair's size (no compression). 
The figure shows that the overall space cost of \idxname is higher than that of other index structures due to \idxname reserving gaps for the key to be inserted. 

However, when the workload is skewed, the gap utilization in cold nodes is low, and thus, it can be compressed to reduce space overhead.
Table 3 in \autoref{fig:figure 12} demonstrates the size of \idxname before and after compressing cold nodes under the Hot-read-A workload. 
The results show that the compression scheme in \idxname can reduce space cost considerably, with compression ratios of 31\% and 37\% for cold nodes on OSM and GENOME, respectively, evaluated using the Hot-write workload.

\textbf{Insight 6:
Compression scheme in \idxname can save space dramatically while ensuring that all gaps are reserved at the hot node, i.e., the gaps are reserved at a more accurate and efficient location than other learned indexes.}

% \begin{table}[hb]% h asks to places the floating element [h]ere.
% \setlength{\abovecaptionskip}{0pt}%    
% \setlength{\belowcaptionskip}{10pt}%
%   \caption{Compress the Cold Node Space}
%   \label{tab:table 3}
%   \begin{tabular}{cccc}
%     \toprule
%     Datasets              &  OSM       & GENOME\\
%     \midrule
%     \idxname size             & 52,293 MB  & 51,016 MB   \\
%     Comp. \idxname size       & 32,311 MB  & 36,166 MB   \\
%     Comp. rate            &    37\%    &     31\%    \\
   
%   \bottomrule
% \end{tabular}
% \end{table}

\vspace{-0.5em}
\subsection{Ablation Study}
\label{Ablation study}

Anneser et al.~\cite{anneser2022hybird} proposed a low-cost sampling method to identify hot and cold data, compress cold data, and expand hot data based on traditional indexes. 
However, due to the need to consider the linear fitting CDF problem, the compression and expansion operations proposed by Anneser et al. cannot be directly applied to the learned index. 
Nonetheless, this low-cost sampling method can be used in \idxname for comparison with our proposed probabilistic framework and high-contend statistics maintenance.

\begin{figure}
\vspace{-1.5em}
% \hspace{-3em}
  \centering
  % \includesvg[width=\linewidth]{figure_14.svg}
  \includegraphics[width=\linewidth]{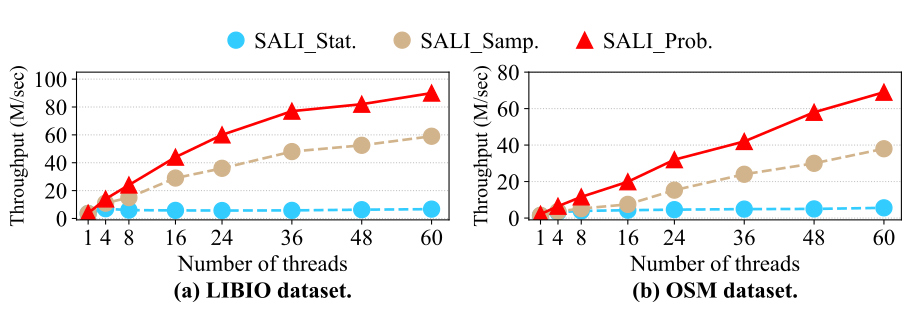}
  \vspace{-2.7em}
  \caption{Comparison of different maintaining statistics methods.}
  \vspace{-2.3em}
  \label{fig:figure 14}
\end{figure}

To compare these methods, we maintain statistical information using three approaches, as shown in ~\autoref{fig:figure 14}: 
\idxname\_Stat. maintains statistical information with the high-contend approach that is employed in many state-of-the-art learned index structures, where every new data insertion triggers a $num.$++ operation in all nodes. 
\idxname\_Samp. maintains statistical information using the sampling method proposed by Anneser et al., where the $num.$++ operation is triggered for every ten inserted data. 
\idxname\_Prob. maintains statistical information using the probability model proposed in our paper.
The results show that \idxname\_Samp. improves scalability compared to \idxname\_Stat. 
Furthermore, the decentralized probability-based model \idxname\_Prob. is more scalable than \idxname\_Samp., achieving up to 35\% higher performance at 60 threads.

\textbf{Insight 7: 
Probability-based models exhibit excellent performance in concurrent scenarios owing to lightweight statistical information maintenance. 
In contrast, any centralized maintenance of statistics can result in performance loss.}

% \vspace{-1em}

% \vspace{-0.7em}
\section{Discussion}
\label{discussion}
% \vspace{-0.3em}

\subsection{Generalizability and Applicability}

We will describe the two main ideas in the \idxname framework, the node-evolving strategies (Section \ref{Evolving strategies}) and the probability model for node statistics (Section \ref{Probability Model}), as general approaches that can potentially be applied to a wide range of learned indexes.

1) What we want to emphasize is that in any application scenario, including but not limited to concurrent settings, if maintaining statistics information globally becomes a performance bottleneck for the index, adopting a probability-based lightweight statistical maintenance approach can help enhance performance.
For example, Lan et al.~\cite{lan2023disklearnedindex} evaluated the performance of learned index approaches on disk and mentioned that the maintenance statistics overhead in both ALEX and LIPP can hurt overall performance because fetching more blocks is required during statistics maintenance. Therefore, our lightweight probability models can address this bottleneck.
Anneser et al.~\cite{anneser2022hybird} still use a global approach to maintain hot-node information in traditional indexes. Although they designed a sampling method to reduce overhead, our proposed probability models have minimal overhead (see ~\autoref{fig:figure 14}). Therefore, using probability models for maintaining hot-node information has the potential to improve performance.
Li et al.~\cite{li2023dili} designed a new model-based learned index framework named DILI, which combines the structures of ALEX and LIPP. The probability models can also serve this framework to improve its scalability.

2) The node-evolving strategies can be applied to different learned indexes. The read evolving can reduce the height of the hot sub-tree, improving read performance.
The write evolving can allocate more space for write-hot nodes in both buffer-based and model-based learned indexes, thereby enhancing insertion performance.
The cold node-compression strategy can save space overhead in model-based learned indexes.
Moreover, if applied to buffer-based learned indexes, new compression algorithms need to be designed to reduce space overhead, which would be an interesting research direction.
Furthermore, Numerous exceptional compression works demonstrate that data compression is a promising research direction~\cite{zhang2016reducing,zhang2020order}. 
Theoretically, the \idxname framework supports the implementation of any excellent compression algorithm. 
Improving upon current compression strategies is a key focus area for future.

3) The current experimental results show that the read-evolving strategy of \idxname is effective only when a slot connects two nodes in the flattened structure. 
As described in Section \ref{Evolving triggered by hot read}, the reason is that the read performance improvement from flattening cannot cover the overhead of using SIMD to accelerate the node lookup process.
This is because the flattened hot subtree contains a relatively small number of keys, and two nodes are sufficient to meet the flattening requirement. 
The number of keys in the subtree depends on the size of the experimental dataset and the LIPP structure.
However, in scenarios with lookup large data volumes, such as Hybrid Transaction/Analytical Processing (HTAP), the read-evolving strategy of the index requires more nodes to be flattened in the same slot and accelerated by SIMD for the lookup process, resulting in performance benefits.
In future work, we plan to design an automatic mechanism to determine the optimal number of nodes to be flattened based on the current data volume and distribution.

%3）目前实验结果表明，SALI的read-evolving策略在hard数据集中才有效。并且扁平化结构中，一个slot连接两个node的情况下有效，其它个数的node实验结果请参见附录。正如section 4.3.1中所述，原因是因为需要扁平化的hot子树中key的数量较少导致随着node增加,扁平化提升的性能收益无法cover住SIMD查找key的开销从而导致性能下降。子树中key的数量较少与实验数据集大小以及SALI所采用的LIPP结构有关。但我们认为在 Hybrid Transaction/Analytical Processing (HTAP)的较大数据量的场景中，索引read-evolving策略需要更多的node被扁平化在同一slot下并通过SIMD加速查找过程，从而得到性能收益。我们拟将在未来工作中设计read-evolving自动判定适合当前数据量及数据分布的最佳扁平化的node的数量。

\vspace{-0.5em}
\subsection{Limitation}

We have identified several limitations in this study:

1) The read-evolving strategy offers significant advantages in scenarios with complex data distributions. Insight 4 in Section \ref{Evolving triggered by hot read} states that flattened tree structures under easy datasets do not evolve. In such cases, enabling the read-evolving strategy would introduce additional overhead to determine if a target node is a hot node, which would compromise the read performance of the \idxname. The overhead of node determination is detailed in additional experiments in the appendix~\cite{appendix}.
Therefore, we have encapsulated read-evolving as a toggle switch to be enabled in scenarios where it is needed. 
However, we acknowledge that determining the benefit of enabling read-evolving in a specific data distribution can be challenging. This will be a focus of our future work.

2) Currently, \idxname does not support duplicate keys. 
The reason for this is that SALI, being based on the LIPP structure, does not currently support the insertion of duplicate data. 
However, Wu et al.~\cite{wu2021updatable} have suggested that it is relatively straightforward for indexes to accommodate duplicate keys, such as by maintaining a pointer to an overflow list. As part of our future work, we plan to focus on implementing the insertion of duplicate data.

\vspace{-0.5em}
\section{Related work}
\label{related work}

In 2018, Kraska et al.\cite{kraska2018case, marcus2020benchmarking, binnig2018tree} introduced a learned index called RMI, which sparked a new wave of index design considerations. 
While the lookup performance was satisfactory, the first-generation learned indexes, such as RMI and RS\cite{kipf2020radixspline}, did not support updates. 
To address this limitation, Galakat et al.\cite{galakatos2019fiting} designed FITing-tree, an updatable learned index.
Ferragina et al.~\cite{ferragina2020pgm} improved the construction algorithm of FITing-tree and introduced the PGM index, which optimized the number of linear models generated while setting the maximum error and used an insertion strategy similar to LSM-Tree~\cite{o1996lsm-tree} to ensure worst-case insertion performance. 
However, FITing-tree and PGM suffered from significant lookup errors and had no better insertion performance than traditional indexes due to their buffer-based insertion strategy. 
In response, Ding et al.\cite{ding2020alex} designed ALEX, which raised the insertion performance of learned indexes to a new level by using a model-based insertion strategy. 
However, ALEX had prediction errors and coarse-grained write lock due to its "shift" strategy to resolve conflicts. 
Wu et al.\cite{wu2021updatable} designed LIPP, another learned index with an error-free model-based insertion strategy. However, LIPP's high-contend statistics maintenance approach in every node hindered scalability.
TONE~\cite{zhang2022tone} mitigates tail latency by dynamically allocating a secondary array to accommodate data, building upon the foundation of ALEX.

XIndex~\cite{tang2020xindex,wang2022concurrent} was the first to implement a concurrent update-capable learned index. 
However, frequent ``last mile'' queries made it less competitive. 
FINEdex~\cite{li2021finedex} improved the concurrency performance by using a flattened structure to avoid coarse-grained locking. 
Wongkham et al.~\cite{gre} implemented concurrent structures for ALEX and LIPP, named ALEX+ and LIPP+, respectively. 
Experimental showed that ALEX+ outperformed LIPP+.

Learned indexes have also inspired new design ideas for other application scenarios~\cite{zhang2022carmi,yu2023lifoss,li2020lisa,ding2020tsunami,dai2020wisckey,maltry2022critical,kipf2019sosd,kraska2018case,marcus2019neo}. 
For instance, Lu et al.~\cite{lu2021apex} developed APEX, a learned index based on NVM. Ma et al.~\cite{ma2022film} designed FILM, a learned index that supports larger-than-memory databases. 
Wu et al.~\cite{wu2022nfl} introduced NFL, a learned index that changes the CDF of stored data through deep learning, making it easier to approximate. 
Nathan et al.~\cite{nathan2020learning} focused on multi-dimensional in-memory learned indexes.
However, these works are beyond the scope of our discussion.

\vspace{-0.5em}
\section{Conclusion}
\label{conclusion}

We have developed \idxname, a highly scalable learned index framework.
In \idxname, we have designed a probability-based framework for monitoring the ``degradation signals'' of the index and identifying hot/cold nodes in a decentralized manner, thereby eliminating thread blocking and improving the index's scalability in a concurrent scenario. 
Since the statistical overhead is negligible, the probability framework provides the necessary conditions for the index to evolve separately toward hot and cold data. 
Furthermore, we have devised evolution strategies that allow \idxname to develop into better-performing local structures for hot and cold nodes independently. 
The experimental results demonstrate that \idxname built upon the $Mod.+C$ structure offers significantly better scalability than state-of-the-art learned indexes, and the evolution strategies can increase read and write performance by at least 25\% and 30\%, respectively.

\balance
%致谢
\vspace{-0.5em}
\begin{acks}
This work is supported by National Natural Science Foundation of China (No. 61972402 and 61972275). 
The corresponding author is Yunpeng Chai (ypchai@ruc.edu.cn).
\end{acks}

\bibliographystyle{ACM-Reference-Format}
\bibliography{0.SALI}


\begin{thebibliography}{45}

%%% the Web address.

\ifx \showCODEN    \undefined \def \showCODEN     #1{\unskip}     \fi
\ifx \showDOI      \undefined \def \showDOI       #1{#1}\fi
\ifx \showISBNx    \undefined \def \showISBNx     #1{\unskip}     \fi
\ifx \showISBNxiii \undefined \def \showISBNxiii  #1{\unskip}     \fi
\ifx \showISSN     \undefined \def \showISSN      #1{\unskip}     \fi
\ifx \showLCCN     \undefined \def \showLCCN      #1{\unskip}     \fi
\ifx \shownote     \undefined \def \shownote      #1{#1}          \fi
\ifx \showarticletitle \undefined \def \showarticletitle #1{#1}   \fi
\ifx \showURL      \undefined \def \showURL       {\relax}        \fi
% invisible to TeX
\providecommand\bibfield[2]{#2}
\providecommand\bibinfo[2]{#2}
\providecommand\natexlab[1]{#1}
\providecommand\showeprint[2][]{arXiv:#2}

\bibitem[Anneser et~al\mbox{.}(2022)]%
        {anneser2022hybird}
\bibfield{author}{\bibinfo{person}{Christoph Anneser}, \bibinfo{person}{Andreas
  Kipf}, \bibinfo{person}{Huanchen Zhang}, \bibinfo{person}{Thomas Neumann},
  {and} \bibinfo{person}{Alfons Kemper}.} \bibinfo{year}{2022}\natexlab{}.
\newblock \showarticletitle{Adaptive Hybrid Indexes}. In
  \bibinfo{booktitle}{\emph{Proceedings of the 2022 International Conference on
  Management of Data}}. \bibinfo{pages}{1626--1639}.
\newblock


\bibitem[Dai et~al\mbox{.}(2020)]%
        {dai2020wisckey}
\bibfield{author}{\bibinfo{person}{Yifan Dai}, \bibinfo{person}{Yien Xu},
  \bibinfo{person}{Aishwarya Ganesan}, \bibinfo{person}{Ramnatthan Alagappan},
  \bibinfo{person}{Brian Kroth}, \bibinfo{person}{Andrea~C Arpaci-Dusseau},
  {and} \bibinfo{person}{Remzi~H Arpaci-Dusseau}.}
  \bibinfo{year}{2020}\natexlab{}.
\newblock \showarticletitle{From wisckey to bourbon: A learned index for
  log-structured merge trees}. In \bibinfo{booktitle}{\emph{Proceedings of the
  14th USENIX Conference on Operating Systems Design and Implementation}}.
  \bibinfo{pages}{155--171}.
\newblock


\bibitem[Ding et~al\mbox{.}(2020a)]%
        {ding2020alex}
\bibfield{author}{\bibinfo{person}{Jialin Ding}, \bibinfo{person}{Umar~Farooq
  Minhas}, \bibinfo{person}{Jia Yu}, \bibinfo{person}{Chi Wang},
  \bibinfo{person}{Jaeyoung Do}, \bibinfo{person}{Yinan Li},
  \bibinfo{person}{Hantian Zhang}, \bibinfo{person}{Badrish Chandramouli},
  \bibinfo{person}{Johannes Gehrke}, \bibinfo{person}{Donald Kossmann},
  {et~al\mbox{.}}} \bibinfo{year}{2020}\natexlab{a}.
\newblock \showarticletitle{ALEX: an updatable adaptive learned index}. In
  \bibinfo{booktitle}{\emph{Proceedings of the 2020 ACM SIGMOD International
  Conference on Management of Data}}. \bibinfo{pages}{969--984}.
\newblock


\bibitem[Ding et~al\mbox{.}(2020b)]%
        {ding2020tsunami}
\bibfield{author}{\bibinfo{person}{Jialin Ding}, \bibinfo{person}{Vikram
  Nathan}, \bibinfo{person}{Mohammad Alizadeh}, {and} \bibinfo{person}{Tim
  Kraska}.} \bibinfo{year}{2020}\natexlab{b}.
\newblock \showarticletitle{Tsunami: a learned multi-dimensional index for
  correlated data and skewed workloads}.
\newblock \bibinfo{journal}{\emph{Proceedings of the VLDB Endowment}}
  \bibinfo{volume}{14}, \bibinfo{number}{2} (\bibinfo{year}{2020}),
  \bibinfo{pages}{74--86}.
\newblock


\bibitem[Ferragina and Vinciguerra(2020)]%
        {ferragina2020pgm}
\bibfield{author}{\bibinfo{person}{Paolo Ferragina} {and}
  \bibinfo{person}{Giorgio Vinciguerra}.} \bibinfo{year}{2020}\natexlab{}.
\newblock \showarticletitle{The PGM-index: a fully-dynamic compressed learned
  index with provable worst-case bounds}.
\newblock \bibinfo{journal}{\emph{Proceedings of the VLDB Endowment}}
  \bibinfo{volume}{13}, \bibinfo{number}{8} (\bibinfo{year}{2020}),
  \bibinfo{pages}{1162--1175}.
\newblock


\bibitem[Fraser(2004)]%
        {fraser2004practical}
\bibfield{author}{\bibinfo{person}{K Fraser}.} \bibinfo{year}{2004}\natexlab{}.
\newblock \showarticletitle{Practical lock-freedom (Doctoral dissertation,
  University of Cambridge)}.
\newblock  (\bibinfo{year}{2004}).
\newblock


\bibitem[Galakatos et~al\mbox{.}(2018)]%
        {binnig2018tree}
\bibfield{author}{\bibinfo{person}{Alex Galakatos}, \bibinfo{person}{Michael
  Markovitch}, \bibinfo{person}{Carsten Binnig}, \bibinfo{person}{Rodrigo
  Fonseca}, {and} \bibinfo{person}{Tim Kraska}.}
  \bibinfo{year}{2018}\natexlab{}.
\newblock \showarticletitle{A-Tree: A Bounded Approximate Index Structure}.
\newblock  (\bibinfo{year}{2018}).
\newblock


\bibitem[Galakatos et~al\mbox{.}(2019)]%
        {galakatos2019fiting}
\bibfield{author}{\bibinfo{person}{Alex Galakatos}, \bibinfo{person}{Michael
  Markovitch}, \bibinfo{person}{Carsten Binnig}, \bibinfo{person}{Rodrigo
  Fonseca}, {and} \bibinfo{person}{Tim Kraska}.}
  \bibinfo{year}{2019}\natexlab{}.
\newblock \showarticletitle{Fiting-tree: A data-aware index structure}. In
  \bibinfo{booktitle}{\emph{Proceedings of the 2019 International Conference on
  Management of Data}}. \bibinfo{pages}{1189--1206}.
\newblock


\bibitem[Ge(2023)]%
        {appendix}
\bibfield{author}{\bibinfo{person}{Jiake Ge}.} \bibinfo{year}{2023}\natexlab{}.
\newblock \bibinfo{title}{Appendix}.
\newblock
\newblock
\urldef\tempurl%
\url{https://github.com/YunWorkshop/SALI/blob/main/SALI\_appendix.pdf}
\showURL{%
\tempurl}


\bibitem[Ge et~al\mbox{.}(2023)]%
        {ge2023learnedindexevaluation}
\bibfield{author}{\bibinfo{person}{Jiake Ge}, \bibinfo{person}{Boyu Shi},
  \bibinfo{person}{Yanfeng Chai}, \bibinfo{person}{Yuanhui Luo},
  \bibinfo{person}{Yunda Guo}, \bibinfo{person}{Yinxuan He}, {and}
  \bibinfo{person}{Yunpeng Chai}.} \bibinfo{year}{2023}\natexlab{}.
\newblock \showarticletitle{Cutting Learned Index into Pieces: An In-depth
  Inquiry into Updatable Learned Indexes}. In \bibinfo{booktitle}{\emph{2023
  IEEE 39th International Conference on Data Engineering (ICDE)}}. IEEE,
  \bibinfo{pages}{315--327}.
\newblock


\bibitem[Gjoka et~al\mbox{.}(2010)]%
        {facebook}
\bibfield{author}{\bibinfo{person}{Minas Gjoka}, \bibinfo{person}{Maciej
  Kurant}, \bibinfo{person}{Carter~T Butts}, {and} \bibinfo{person}{Athina
  Markopoulou}.} \bibinfo{year}{2010}\natexlab{}.
\newblock \showarticletitle{Walking in facebook: A case study of unbiased
  sampling of osns}. In \bibinfo{booktitle}{\emph{2010 Proceedings IEEE
  Infocom}}. IEEE, \bibinfo{pages}{1--9}.
\newblock


\bibitem[Kim et~al\mbox{.}(2010)]%
        {kim2010fast}
\bibfield{author}{\bibinfo{person}{Changkyu Kim}, \bibinfo{person}{Jatin
  Chhugani}, \bibinfo{person}{Nadathur Satish}, \bibinfo{person}{Eric Sedlar},
  \bibinfo{person}{Anthony~D Nguyen}, \bibinfo{person}{Tim Kaldewey},
  \bibinfo{person}{Victor~W Lee}, \bibinfo{person}{Scott~A Brandt}, {and}
  \bibinfo{person}{Pradeep Dubey}.} \bibinfo{year}{2010}\natexlab{}.
\newblock \showarticletitle{FAST: fast architecture sensitive tree search on
  modern CPUs and GPUs}. In \bibinfo{booktitle}{\emph{Proceedings of the 2010
  ACM SIGMOD International Conference on Management of data}}.
  \bibinfo{pages}{339--350}.
\newblock


\bibitem[Kipf et~al\mbox{.}(2019)]%
        {kipf2019sosd}
\bibfield{author}{\bibinfo{person}{Andreas Kipf}, \bibinfo{person}{Ryan
  Marcus}, \bibinfo{person}{Alexander van Renen}, \bibinfo{person}{Mihail
  Stoian}, \bibinfo{person}{Alfons Kemper}, \bibinfo{person}{Tim Kraska}, {and}
  \bibinfo{person}{Thomas Neumann}.} \bibinfo{year}{2019}\natexlab{}.
\newblock \showarticletitle{SOSD: A benchmark for learned indexes}.
\newblock \bibinfo{journal}{\emph{NeurIPS Workshop on Learned Systems}}
  (\bibinfo{year}{2019}).
\newblock


\bibitem[Kipf et~al\mbox{.}(2020)]%
        {kipf2020radixspline}
\bibfield{author}{\bibinfo{person}{Andreas Kipf}, \bibinfo{person}{Ryan
  Marcus}, \bibinfo{person}{Alexander van Renen}, \bibinfo{person}{Mihail
  Stoian}, \bibinfo{person}{Alfons Kemper}, \bibinfo{person}{Tim Kraska}, {and}
  \bibinfo{person}{Thomas Neumann}.} \bibinfo{year}{2020}\natexlab{}.
\newblock \showarticletitle{RadixSpline: a single-pass learned index}. In
  \bibinfo{booktitle}{\emph{Proceedings of the Third International Workshop on
  Exploiting Artificial Intelligence Techniques for Data Management}}.
  \bibinfo{pages}{1--5}.
\newblock


\bibitem[Kraska et~al\mbox{.}(2018)]%
        {kraska2018case}
\bibfield{author}{\bibinfo{person}{Tim Kraska}, \bibinfo{person}{Alex Beutel},
  \bibinfo{person}{Ed~H Chi}, \bibinfo{person}{Jeffrey Dean}, {and}
  \bibinfo{person}{Neoklis Polyzotis}.} \bibinfo{year}{2018}\natexlab{}.
\newblock \showarticletitle{The case for learned index structures}. In
  \bibinfo{booktitle}{\emph{Proceedings of the 2018 international conference on
  management of data}}. \bibinfo{pages}{489--504}.
\newblock


\bibitem[Lan et~al\mbox{.}(2023)]%
        {lan2023disklearnedindex}
\bibfield{author}{\bibinfo{person}{Hai Lan}, \bibinfo{person}{Zhifeng Bao},
  \bibinfo{person}{J~Shane Culpepper}, {and} \bibinfo{person}{Renata
  Borovica-Gajic}.} \bibinfo{year}{2023}\natexlab{}.
\newblock \showarticletitle{Updatable Learned Indexes Meet Disk-Resident
  DBMS-From Evaluations to Design Choices}.
\newblock \bibinfo{journal}{\emph{Proceedings of the ACM on Management of
  Data}} \bibinfo{volume}{1}, \bibinfo{number}{2} (\bibinfo{year}{2023}),
  \bibinfo{pages}{1--22}.
\newblock


\bibitem[Leis et~al\mbox{.}(2018)]%
        {leis2018leanstore}
\bibfield{author}{\bibinfo{person}{Viktor Leis}, \bibinfo{person}{Michael
  Haubenschild}, \bibinfo{person}{Alfons Kemper}, {and} \bibinfo{person}{Thomas
  Neumann}.} \bibinfo{year}{2018}\natexlab{}.
\newblock \showarticletitle{LeanStore: In-memory data management beyond main
  memory}. In \bibinfo{booktitle}{\emph{2018 IEEE 34th International Conference
  on Data Engineering (ICDE)}}. IEEE, \bibinfo{pages}{185--196}.
\newblock


\bibitem[Leis et~al\mbox{.}(2013)]%
        {art2013adaptive}
\bibfield{author}{\bibinfo{person}{Viktor Leis}, \bibinfo{person}{Alfons
  Kemper}, {and} \bibinfo{person}{Thomas Neumann}.}
  \bibinfo{year}{2013}\natexlab{}.
\newblock \showarticletitle{The adaptive radix tree: ARTful indexing for
  main-memory databases}. In \bibinfo{booktitle}{\emph{2013 IEEE 29th
  International Conference on Data Engineering (ICDE)}}. IEEE,
  \bibinfo{pages}{38--49}.
\newblock


\bibitem[Leis et~al\mbox{.}(2016)]%
        {2016artolc}
\bibfield{author}{\bibinfo{person}{V. Leis}, \bibinfo{person}{F. Scheibner},
  \bibinfo{person}{Alfons~Heinrich Kemper}, {and} \bibinfo{person}{T.
  Neumann}.} \bibinfo{year}{2016}\natexlab{}.
\newblock \showarticletitle{The ART of practical synchronization}. In
  \bibinfo{booktitle}{\emph{the 12th International Workshop}}.
\newblock


\bibitem[Levandoski et~al\mbox{.}(2013)]%
        {levandoski2013bw}
\bibfield{author}{\bibinfo{person}{Justin~J Levandoski},
  \bibinfo{person}{David~B Lomet}, {and} \bibinfo{person}{Sudipta Sengupta}.}
  \bibinfo{year}{2013}\natexlab{}.
\newblock \showarticletitle{The Bw-Tree: A B-tree for new hardware platforms}.
  In \bibinfo{booktitle}{\emph{2013 IEEE 29th International Conference on Data
  Engineering (ICDE)}}. IEEE, \bibinfo{pages}{302--313}.
\newblock


\bibitem[Li et~al\mbox{.}(2021)]%
        {li2021finedex}
\bibfield{author}{\bibinfo{person}{Pengfei Li}, \bibinfo{person}{Yu Hua},
  \bibinfo{person}{Jingnan Jia}, {and} \bibinfo{person}{Pengfei Zuo}.}
  \bibinfo{year}{2021}\natexlab{}.
\newblock \showarticletitle{FINEdex: a fine-grained learned index scheme for
  scalable and concurrent memory systems}.
\newblock \bibinfo{journal}{\emph{Proceedings of the VLDB Endowment}}
  \bibinfo{volume}{15}, \bibinfo{number}{2} (\bibinfo{year}{2021}),
  \bibinfo{pages}{321--334}.
\newblock


\bibitem[Li et~al\mbox{.}(2020)]%
        {li2020lisa}
\bibfield{author}{\bibinfo{person}{Pengfei Li}, \bibinfo{person}{Hua Lu},
  \bibinfo{person}{Qian Zheng}, \bibinfo{person}{Long Yang}, {and}
  \bibinfo{person}{Gang Pan}.} \bibinfo{year}{2020}\natexlab{}.
\newblock \showarticletitle{LISA: A learned index structure for spatial data}.
  In \bibinfo{booktitle}{\emph{Proceedings of the 2020 ACM SIGMOD international
  conference on management of data}}. \bibinfo{pages}{2119--2133}.
\newblock


\bibitem[Li et~al\mbox{.}(2023)]%
        {li2023dili}
\bibfield{author}{\bibinfo{person}{Pengfei Li}, \bibinfo{person}{Hua Lu},
  \bibinfo{person}{Rong Zhu}, \bibinfo{person}{Bolin Ding},
  \bibinfo{person}{Long Yang}, {and} \bibinfo{person}{Gang Pan}.}
  \bibinfo{year}{2023}\natexlab{}.
\newblock \showarticletitle{DILI: A Distribution-Driven Learned Index}.
\newblock \bibinfo{journal}{\emph{arXiv preprint arXiv:2304.08817}}
  (\bibinfo{year}{2023}).
\newblock


\bibitem[Lopez and Gallemore(2021)]%
        {lopez2021covid}
\bibfield{author}{\bibinfo{person}{Christian~E Lopez} {and}
  \bibinfo{person}{Caleb Gallemore}.} \bibinfo{year}{2021}\natexlab{}.
\newblock \showarticletitle{An augmented multilingual Twitter dataset for
  studying the COVID-19 infodemic}.
\newblock \bibinfo{journal}{\emph{Social Network Analysis and Mining}}
  \bibinfo{volume}{11}, \bibinfo{number}{1} (\bibinfo{year}{2021}),
  \bibinfo{pages}{102}.
\newblock


\bibitem[Lu et~al\mbox{.}(2021)]%
        {lu2021apex}
\bibfield{author}{\bibinfo{person}{Baotong Lu}, \bibinfo{person}{Jialin Ding},
  \bibinfo{person}{Eric Lo}, \bibinfo{person}{Umar~Farooq Minhas}, {and}
  \bibinfo{person}{Tianzheng Wang}.} \bibinfo{year}{2021}\natexlab{}.
\newblock \showarticletitle{APEX: a high-performance learned index on
  persistent memory}.
\newblock \bibinfo{journal}{\emph{Proceedings of the VLDB Endowment}}
  \bibinfo{volume}{15}, \bibinfo{number}{3} (\bibinfo{year}{2021}),
  \bibinfo{pages}{597--610}.
\newblock


\bibitem[Ma et~al\mbox{.}(2022)]%
        {ma2022film}
\bibfield{author}{\bibinfo{person}{Chaohong Ma}, \bibinfo{person}{Xiaohui Yu},
  \bibinfo{person}{Yifan Li}, \bibinfo{person}{Xiaofeng Meng}, {and}
  \bibinfo{person}{Aishan Maoliniyazi}.} \bibinfo{year}{2022}\natexlab{}.
\newblock \showarticletitle{FILM: A Fully Learned Index for Larger-Than-Memory
  Databases}.
\newblock \bibinfo{journal}{\emph{Proceedings of the VLDB Endowment}}
  \bibinfo{volume}{16}, \bibinfo{number}{3} (\bibinfo{year}{2022}),
  \bibinfo{pages}{561--573}.
\newblock


\bibitem[Maltry and Dittrich(2022)]%
        {maltry2022critical}
\bibfield{author}{\bibinfo{person}{Marcel Maltry} {and} \bibinfo{person}{Jens
  Dittrich}.} \bibinfo{year}{2022}\natexlab{}.
\newblock \showarticletitle{A critical analysis of recursive model indexes}.
\newblock \bibinfo{journal}{\emph{Proceedings of the VLDB Endowment}}
  \bibinfo{volume}{15}, \bibinfo{number}{5} (\bibinfo{year}{2022}),
  \bibinfo{pages}{1079--1091}.
\newblock


\bibitem[Mao et~al\mbox{.}(2012)]%
        {masstree}
\bibfield{author}{\bibinfo{person}{Yandong Mao}, \bibinfo{person}{Eddie
  Kohler}, {and} \bibinfo{person}{Robert~Tappan Morris}.}
  \bibinfo{year}{2012}\natexlab{}.
\newblock \showarticletitle{Cache craftiness for fast multicore key-value
  storage}. In \bibinfo{booktitle}{\emph{Proceedings of the 7th ACM european
  conference on Computer Systems}}. \bibinfo{pages}{183--196}.
\newblock


\bibitem[Marcus et~al\mbox{.}(2020)]%
        {marcus2020benchmarking}
\bibfield{author}{\bibinfo{person}{Ryan Marcus}, \bibinfo{person}{Andreas
  Kipf}, \bibinfo{person}{Alexander van Renen}, \bibinfo{person}{Mihail
  Stoian}, \bibinfo{person}{Sanchit Misra}, \bibinfo{person}{Alfons Kemper},
  \bibinfo{person}{Thomas Neumann}, {and} \bibinfo{person}{Tim Kraska}.}
  \bibinfo{year}{2020}\natexlab{}.
\newblock \showarticletitle{Benchmarking learned indexes}.
\newblock \bibinfo{journal}{\emph{Proceedings of the VLDB Endowment}}
  (\bibinfo{year}{2020}).
\newblock


\bibitem[Marcus et~al\mbox{.}(2019)]%
        {marcus2019neo}
\bibfield{author}{\bibinfo{person}{Ryan Marcus}, \bibinfo{person}{Parimarjan
  Negi}, \bibinfo{person}{Hongzi Mao}, \bibinfo{person}{Chi Zhang},
  \bibinfo{person}{Mohammad Alizadeh}, \bibinfo{person}{Tim Kraska},
  \bibinfo{person}{Olga Papaemmanouil}, {and} \bibinfo{person}{Nesime Tatbul}.}
  \bibinfo{year}{2019}\natexlab{}.
\newblock \showarticletitle{Neo: a learned query optimizer}.
\newblock \bibinfo{journal}{\emph{Proceedings of the VLDB Endowment}}
  \bibinfo{volume}{12}, \bibinfo{number}{11} (\bibinfo{year}{2019}),
  \bibinfo{pages}{1705--1718}.
\newblock


\bibitem[McKenney et~al\mbox{.}(2001)]%
        {mckenney2001rcu}
\bibfield{author}{\bibinfo{person}{Paul~E McKenney}, \bibinfo{person}{Jonathan
  Appavoo}, \bibinfo{person}{Andi Kleen}, \bibinfo{person}{Orran Krieger},
  \bibinfo{person}{Rusty Russell}, \bibinfo{person}{Dipankar Sarma}, {and}
  \bibinfo{person}{Maneesh Soni}.} \bibinfo{year}{2001}\natexlab{}.
\newblock \showarticletitle{Read-copy update}. In
  \bibinfo{booktitle}{\emph{AUUG Conference Proceedings}}. AUUG, Inc.,
  \bibinfo{pages}{175}.
\newblock


\bibitem[Nathan et~al\mbox{.}(2020)]%
        {nathan2020learning}
\bibfield{author}{\bibinfo{person}{Vikram Nathan}, \bibinfo{person}{Jialin
  Ding}, \bibinfo{person}{Mohammad Alizadeh}, {and} \bibinfo{person}{Tim
  Kraska}.} \bibinfo{year}{2020}\natexlab{}.
\newblock \showarticletitle{Learning multi-dimensional indexes}. In
  \bibinfo{booktitle}{\emph{Proceedings of the 2020 ACM SIGMOD international
  conference on management of data}}. \bibinfo{pages}{985--1000}.
\newblock


\bibitem[O’Neil et~al\mbox{.}(1996)]%
        {o1996lsm-tree}
\bibfield{author}{\bibinfo{person}{Patrick O’Neil}, \bibinfo{person}{Edward
  Cheng}, \bibinfo{person}{Dieter Gawlick}, {and} \bibinfo{person}{Elizabeth
  O’Neil}.} \bibinfo{year}{1996}\natexlab{}.
\newblock \showarticletitle{The log-structured merge-tree (LSM-tree)}.
\newblock \bibinfo{journal}{\emph{Acta Informatica}} \bibinfo{volume}{33},
  \bibinfo{number}{4} (\bibinfo{year}{1996}), \bibinfo{pages}{351--385}.
\newblock


\bibitem[Rao et~al\mbox{.}(2014)]%
        {rao20143genome}
\bibfield{author}{\bibinfo{person}{Suhas~SP Rao}, \bibinfo{person}{Miriam~H
  Huntley}, \bibinfo{person}{Neva~C Durand}, \bibinfo{person}{Elena~K
  Stamenova}, \bibinfo{person}{Ivan~D Bochkov}, \bibinfo{person}{James~T
  Robinson}, \bibinfo{person}{Adrian~L Sanborn}, \bibinfo{person}{Ido Machol},
  \bibinfo{person}{Arina~D Omer}, \bibinfo{person}{Eric~S Lander},
  {et~al\mbox{.}}} \bibinfo{year}{2014}\natexlab{}.
\newblock \showarticletitle{A 3D map of the human genome at kilobase resolution
  reveals principles of chromatin looping}.
\newblock \bibinfo{journal}{\emph{Cell}} \bibinfo{volume}{159},
  \bibinfo{number}{7} (\bibinfo{year}{2014}), \bibinfo{pages}{1665--1680}.
\newblock


\bibitem[Siakavaras et~al\mbox{.}(2020)]%
        {siakavaras2020efficient}
\bibfield{author}{\bibinfo{person}{Dimitrios Siakavaras},
  \bibinfo{person}{Panagiotis Billis}, \bibinfo{person}{Konstantinos Nikas},
  \bibinfo{person}{Georgios Goumas}, {and} \bibinfo{person}{Nectarios
  Koziris}.} \bibinfo{year}{2020}\natexlab{}.
\newblock \showarticletitle{Efficient Concurrent Range Queries in B+-trees
  using RCU-HTM}. In \bibinfo{booktitle}{\emph{Proceedings of the 32nd ACM
  Symposium on Parallelism in Algorithms and Architectures}}.
  \bibinfo{pages}{571--573}.
\newblock


\bibitem[Tang et~al\mbox{.}(2020)]%
        {tang2020xindex}
\bibfield{author}{\bibinfo{person}{Chuzhe Tang}, \bibinfo{person}{Youyun Wang},
  \bibinfo{person}{Zhiyuan Dong}, \bibinfo{person}{Gansen Hu},
  \bibinfo{person}{Zhaoguo Wang}, \bibinfo{person}{Minjie Wang}, {and}
  \bibinfo{person}{Haibo Chen}.} \bibinfo{year}{2020}\natexlab{}.
\newblock \showarticletitle{XIndex: a scalable learned index for multicore data
  storage}. In \bibinfo{booktitle}{\emph{Proceedings of the 25th ACM SIGPLAN
  Symposium on Principles and Practice of Parallel Programming}}.
  \bibinfo{pages}{308--320}.
\newblock


\bibitem[Wang et~al\mbox{.}(2022)]%
        {wang2022concurrent}
\bibfield{author}{\bibinfo{person}{Zhaoguo Wang}, \bibinfo{person}{Haibo Chen},
  \bibinfo{person}{Youyun Wang}, \bibinfo{person}{Chuzhe Tang}, {and}
  \bibinfo{person}{Huan Wang}.} \bibinfo{year}{2022}\natexlab{}.
\newblock \showarticletitle{The concurrent learned indexes for multicore data
  storage}.
\newblock \bibinfo{journal}{\emph{ACM Transactions on Storage (TOS)}}
  \bibinfo{volume}{18}, \bibinfo{number}{1} (\bibinfo{year}{2022}),
  \bibinfo{pages}{1--35}.
\newblock


\bibitem[Wongkham et~al\mbox{.}(2022)]%
        {gre}
\bibfield{author}{\bibinfo{person}{Chaichon Wongkham}, \bibinfo{person}{Baotong
  Lu}, \bibinfo{person}{Chris Liu}, \bibinfo{person}{Zhicong Zhong},
  \bibinfo{person}{Eric Lo}, {and} \bibinfo{person}{Tianzheng Wang}.}
  \bibinfo{year}{2022}\natexlab{}.
\newblock \showarticletitle{Are Updatable Learned Indexes Ready?}
\newblock \bibinfo{journal}{\emph{Proceedings of the VLDB Endowment}}
  (\bibinfo{year}{2022}).
\newblock


\bibitem[Wu et~al\mbox{.}(2021)]%
        {wu2021updatable}
\bibfield{author}{\bibinfo{person}{Jiacheng Wu}, \bibinfo{person}{Yong Zhang},
  \bibinfo{person}{Shimin Chen}, \bibinfo{person}{Jin Wang},
  \bibinfo{person}{Yu Chen}, {and} \bibinfo{person}{Chunxiao Xing}.}
  \bibinfo{year}{2021}\natexlab{}.
\newblock \showarticletitle{Updatable learned index with precise positions}.
\newblock \bibinfo{journal}{\emph{Proceedings of the VLDB Endowment}}
  \bibinfo{volume}{14}, \bibinfo{number}{8} (\bibinfo{year}{2021}),
  \bibinfo{pages}{1276--1288}.
\newblock


\bibitem[Wu et~al\mbox{.}(2022)]%
        {wu2022nfl}
\bibfield{author}{\bibinfo{person}{Shangyu Wu}, \bibinfo{person}{Yufei Cui},
  \bibinfo{person}{Jinghuan Yu}, \bibinfo{person}{Xuan Sun},
  \bibinfo{person}{Tei-Wei Kuo}, {and} \bibinfo{person}{Chun~Jason Xue}.}
  \bibinfo{year}{2022}\natexlab{}.
\newblock \showarticletitle{NFL: robust learned index via distribution
  transformation}.
\newblock \bibinfo{journal}{\emph{Proceedings of the VLDB Endowment}}
  \bibinfo{volume}{15}, \bibinfo{number}{10} (\bibinfo{year}{2022}),
  \bibinfo{pages}{2188--2200}.
\newblock


\bibitem[Yu et~al\mbox{.}(2023)]%
        {yu2023lifoss}
\bibfield{author}{\bibinfo{person}{Tong Yu}, \bibinfo{person}{Guanfeng Liu},
  \bibinfo{person}{An Liu}, \bibinfo{person}{Zhixu Li}, {and}
  \bibinfo{person}{Lei Zhao}.} \bibinfo{year}{2023}\natexlab{}.
\newblock \showarticletitle{LIFOSS: a learned index scheme for streaming
  scenarios}.
\newblock \bibinfo{journal}{\emph{World Wide Web}} \bibinfo{volume}{26},
  \bibinfo{number}{1} (\bibinfo{year}{2023}), \bibinfo{pages}{501--518}.
\newblock


\bibitem[Zhang et~al\mbox{.}(2016)]%
        {zhang2016reducing}
\bibfield{author}{\bibinfo{person}{Huanchen Zhang}, \bibinfo{person}{David~G
  Andersen}, \bibinfo{person}{Andrew Pavlo}, \bibinfo{person}{Michael
  Kaminsky}, \bibinfo{person}{Lin Ma}, {and} \bibinfo{person}{Rui Shen}.}
  \bibinfo{year}{2016}\natexlab{}.
\newblock \showarticletitle{Reducing the storage overhead of main-memory OLTP
  databases with hybrid indexes}. In \bibinfo{booktitle}{\emph{Proceedings of
  the 2016 International Conference on Management of Data}}.
  \bibinfo{pages}{1567--1581}.
\newblock


\bibitem[Zhang et~al\mbox{.}(2020)]%
        {zhang2020order}
\bibfield{author}{\bibinfo{person}{Huanchen Zhang}, \bibinfo{person}{Xiaoxuan
  Liu}, \bibinfo{person}{David~G Andersen}, \bibinfo{person}{Michael Kaminsky},
  \bibinfo{person}{Kimberly Keeton}, {and} \bibinfo{person}{Andrew Pavlo}.}
  \bibinfo{year}{2020}\natexlab{}.
\newblock \showarticletitle{Order-preserving key compression for in-memory
  search trees}. In \bibinfo{booktitle}{\emph{Proceedings of the 2020 ACM
  SIGMOD International Conference on Management of Data}}.
  \bibinfo{pages}{1601--1615}.
\newblock


\bibitem[Zhang and Gao(2022)]%
        {zhang2022carmi}
\bibfield{author}{\bibinfo{person}{Jiaoyi Zhang} {and} \bibinfo{person}{Yihan
  Gao}.} \bibinfo{year}{2022}\natexlab{}.
\newblock \showarticletitle{CARMI: a cache-aware learned index with a
  cost-based construction algorithm}.
\newblock \bibinfo{journal}{\emph{Proceedings of the VLDB Endowment}}
  \bibinfo{volume}{15}, \bibinfo{number}{11} (\bibinfo{year}{2022}),
  \bibinfo{pages}{2679--2691}.
\newblock


\bibitem[Zhang et~al\mbox{.}(2022)]%
        {zhang2022tone}
\bibfield{author}{\bibinfo{person}{Yong Zhang}, \bibinfo{person}{Xinran Xiong},
  {and} \bibinfo{person}{Oana Balmau}.} \bibinfo{year}{2022}\natexlab{}.
\newblock \showarticletitle{TONE: cutting tail-latency in learned indexes}. In
  \bibinfo{booktitle}{\emph{Proceedings of the Workshop on Challenges and
  Opportunities of Efficient and Performant Storage Systems}}.
  \bibinfo{pages}{16--23}.
\newblock


\end{thebibliography}

%   附录 %%%%%%
% If your work has an appendix, this is the place to put it.
\appendix
\section{Comprehensive evaluation results}

\subsection{Throughput}

\begin{figure*}
\vspace{-1em}
% \hspace{-3em}
  \centering
  % \includesvg[width=1\linewidth]{./appendix_figure_Throughput_write-only.svg}
  \includegraphics[width=1\linewidth]{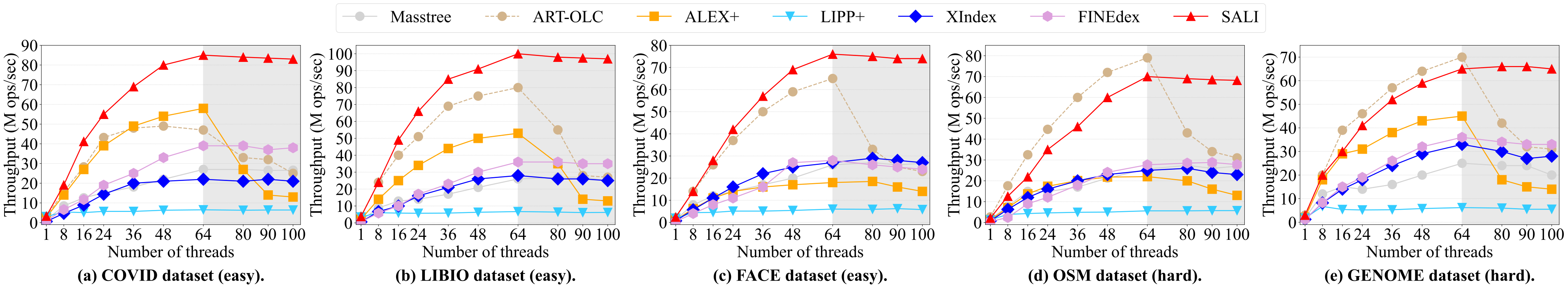}
  \vspace{-2.5em}
  \caption{The indexes scalability on write-only workloads (Throughput).}
  \vspace{-1.5em}
  \label{fig:figure 16}
\end{figure*}

\begin{figure*}
  \centering
  % \includesvg[width=1\linewidth]{./appendix_figure_Throughput_5050.svg}
  \includegraphics[width=1\linewidth]{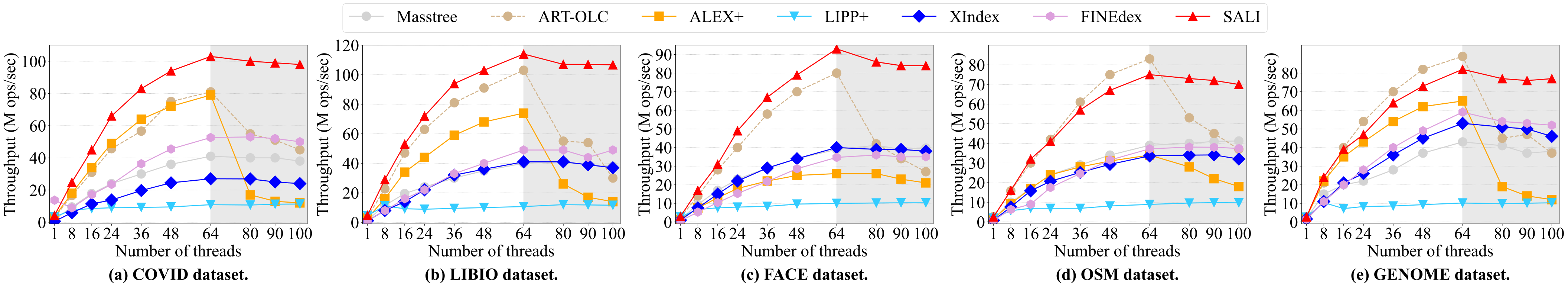}
  \vspace{-2.5em}
  \caption{The indexes scalability on balance workloads (Throughput).}
  \vspace{-1.5em}
  \label{fig:figure 17}
\end{figure*}

\begin{figure*}
% \vspace{-1em}
% \hspace{-3em}
  \centering
  % \includesvg[width=1\linewidth]{./appendix_figure_Throughput_8020.svg}
  \includegraphics[width=1\linewidth]{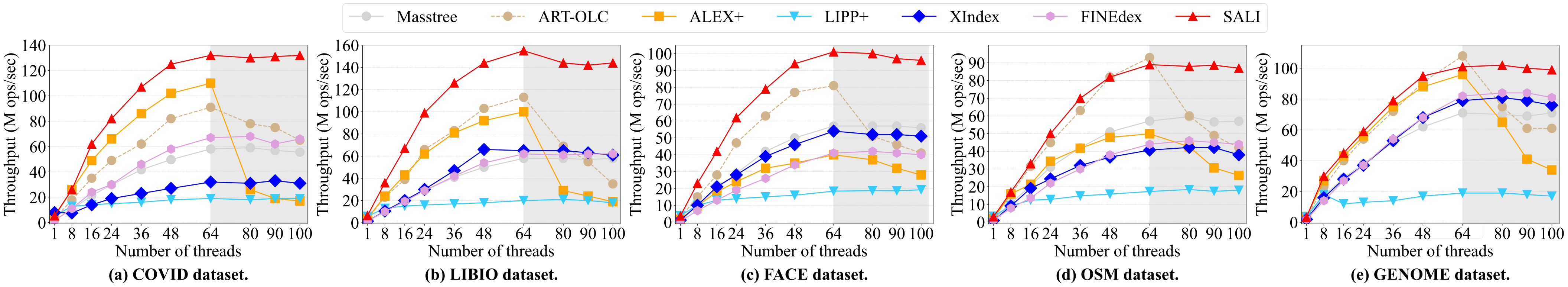}
  \vspace{-2.5em}
  \caption{The indexes scalability on read-intensive workloads (Throughput).}
  \vspace{-1.5em}
  \label{fig:figure 18}
\end{figure*}

\begin{figure*}
% \vspace{-1em}
% \hspace{-3em}
  \centering
  % \includesvg[width=1\linewidth]{./appendix_figure_Throughput_read-only.svg}
  \includegraphics[width=1\linewidth]{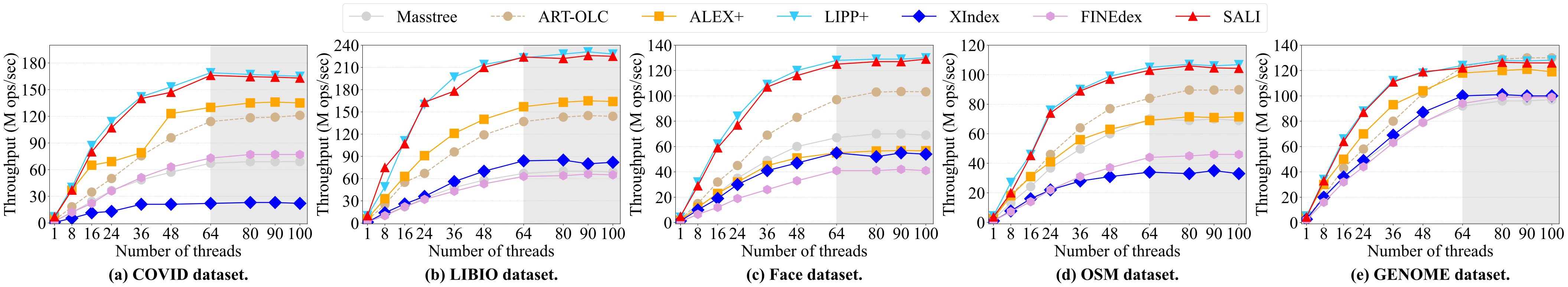}
  \vspace{-2.5em}
  \caption{The indexes scalability on read-only workloads (Throughput).}
  \vspace{-1.5em}
  \label{fig:figure 19}
\end{figure*}

\autoref{fig:figure 16}-\ref{fig:figure 19} showcase the complete evaluation results for all datasets and different read/write ratios (uniform distribution), regarding the complete assessment of throughput as mentioned in the main text. \idxname exhibits superior throughput compared to other SOTA learned indexes.

\subsection{Tail latency}

\begin{figure*}
\vspace{-1em}
% \hspace{-3em}
  \centering
  % \includesvg[width=1\linewidth]{./appendix_figure_latency_write-only.svg}
  \includegraphics[width=1\linewidth]{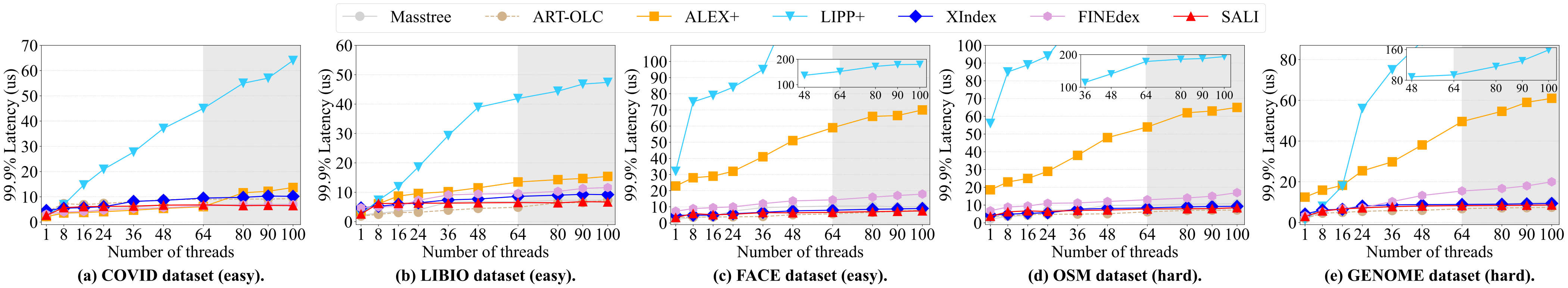}
  \vspace{-2.5em}
  \caption{The indexes scalability on write-only workloads (Latency).}
  % \vspace{-1.5em}
  \label{fig:figure 20}
\end{figure*}

\begin{figure*}
% \hspace{-3em}
  \centering
  % \includesvg[width=1\linewidth]{./appendix_figure_latency_5050.svg}
  \includegraphics[width=1\linewidth]{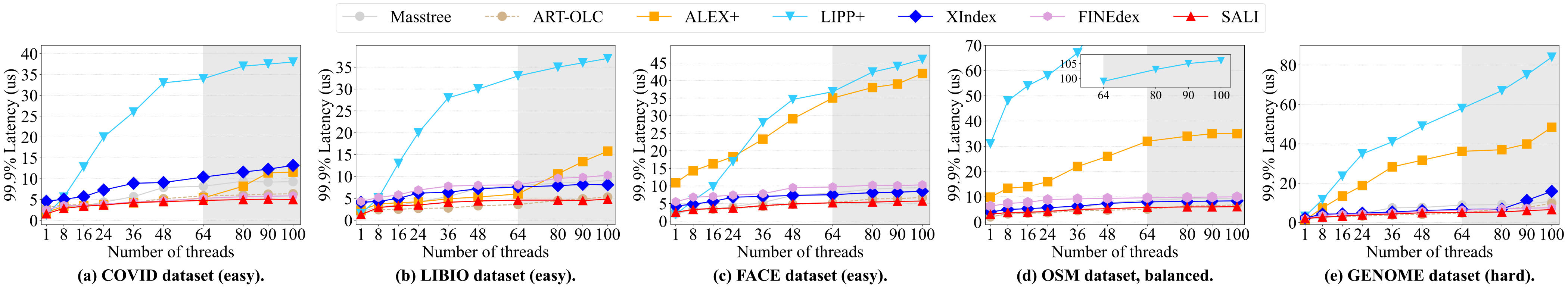}
  \vspace{-2.5em}
  \caption{The indexes scalability on balance workloads (Latency).}
  % \vspace{-1.5em}
  \label{fig:figure 21}
\end{figure*}

\begin{figure*}
% \hspace{-3em}
  \centering
  % \includesvg[width=1\linewidth]{./appendix_figure_latency_8020.svg}
  \includegraphics[width=1\linewidth]{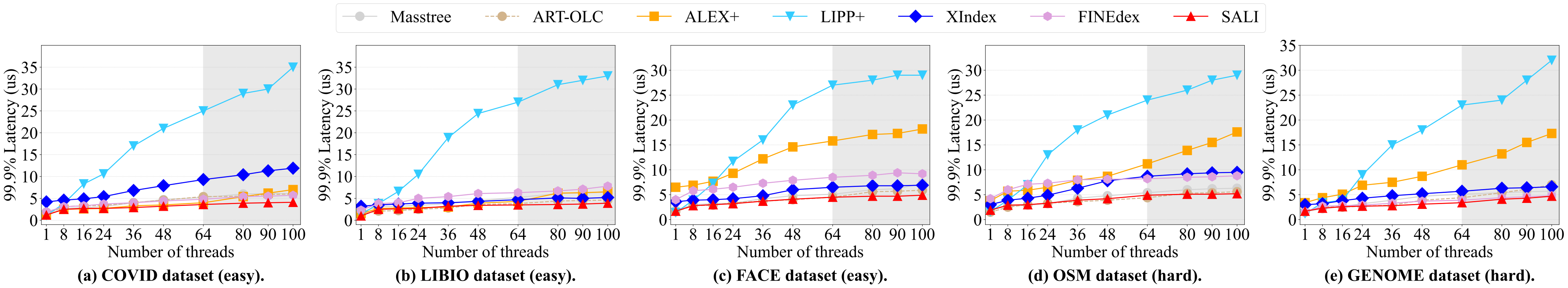}
  \vspace{-2.5em}
  \caption{The indexes scalability on read-intensive workloads (Latency).}
  % \vspace{-1.5em}
  \label{fig:figure 22}
\end{figure*}

\begin{figure*}
\vspace{-1em}
% \hspace{-3em}
  \centering
  % \includesvg[width=1\linewidth]{./appendix_figure_latency_read-only.svg}
  \includegraphics[width=1\linewidth]{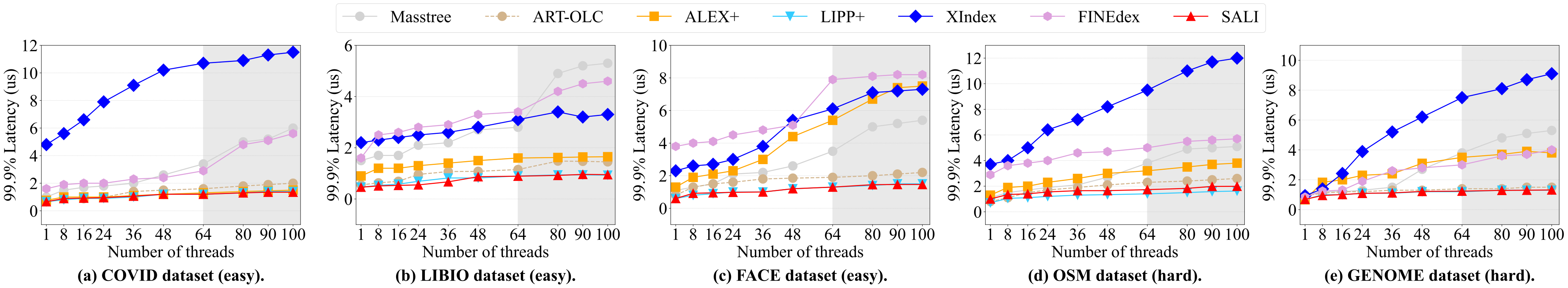}
  \vspace{-2.5em}
  \caption{The indexes scalability on read-only workloads (Latency).}
  \vspace{-1.2em}
  \label{fig:figure 23}
\end{figure*}

\autoref{fig:figure 20}-\ref{fig:figure 23} present the comprehensive evaluation results for all datasets and various read-write ratios (uniform distribution), regarding the complete assessment of tail latency as mentioned in the main text. \idxname maintains low tail latency across the board.

\vspace{-0.5em}
\subsection{Index building and range queries}

\begin{figure}
% \hspace{-3em}
% \vspace{-1.3em}
  \centering
  \includegraphics[width=1\linewidth]{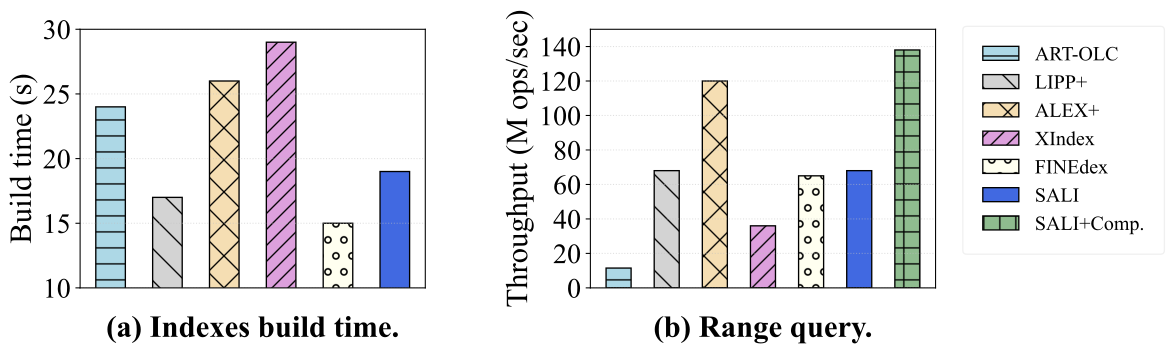}
  \vspace{-2.5em}
  \caption{bulk time and range query performance.}
  \vspace{-1em}
  \label{fig:figure 15}
\end{figure}

\autoref{fig:figure 15}(a) presents the time required for bulk loading 100M keys by various learned indexes. 
Notably, \idxname's build time is significantly lower than that of ALEX+ and XIndex. 
This is because \idxname creates new nodes for conflicting keys and does not require key movement to maintain gaps, as in ALEX+. 
\idxname needs to create a cooling pool, resulting in a slightly longer build time than LIPP+.

\autoref{fig:figure 15}(b) presents the evaluation of a range query of 100 keys with 48 threads. 
\idxname outperforms XIndex and FINEdex but falls short of ALEX+. 
This is because \idxname's node layout, which resembles that of a B-tree, contains more gaps and interleaves child pointers and data in the node array. 
Consequently, scan on the array encounters many branches. 
To mitigate this issue, we make preliminary optimizations by compressing the hot scan node (similar to the cold nodes compression designed in \autoref{fig:figure 5}(c)) to remove the gap and store the data in a node, as illustrated in \idxname+Comp. in \autoref{fig:figure 15}(b), assuming that we know the hot scan information. 
The results show that this approach outperforms other baselines. 
Our future research will focus on designing more robust methods for identifying hot scan nodes, building on the prerequisites provided by the lightweight models.

\begin{figure*}
\vspace{-1em}
% \hspace{-3em}
  \centering
  \includegraphics[width=0.8\linewidth]{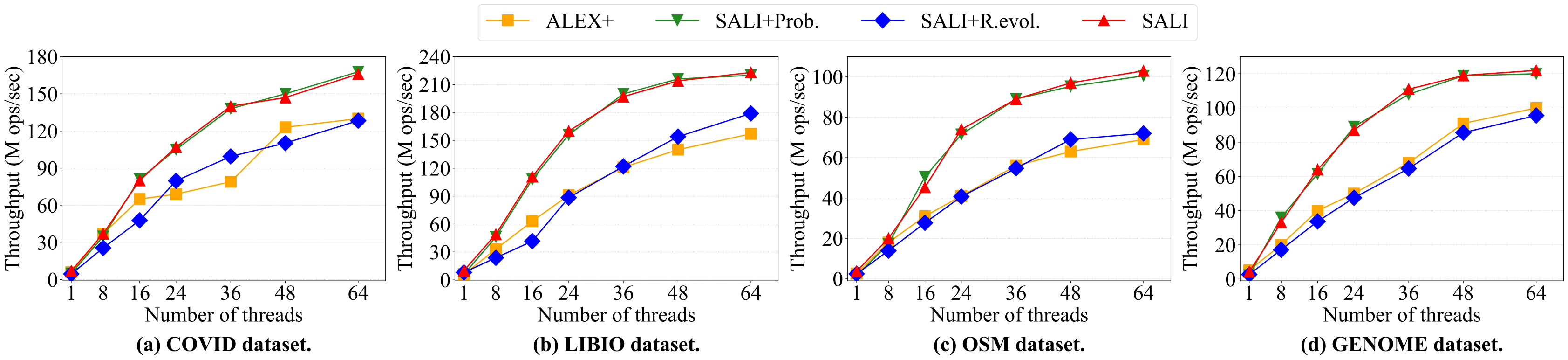}
  \vspace{-1.5em}
  \caption{Overhead analysis of the read-evolving strategy.}
  % \vspace{-1.5em}
  \label{fig:figure 24}
\end{figure*}

\begin{figure*}
\vspace{-1em}
% \hspace{-3em}
  \centering
  \includegraphics[width=0.8\linewidth]{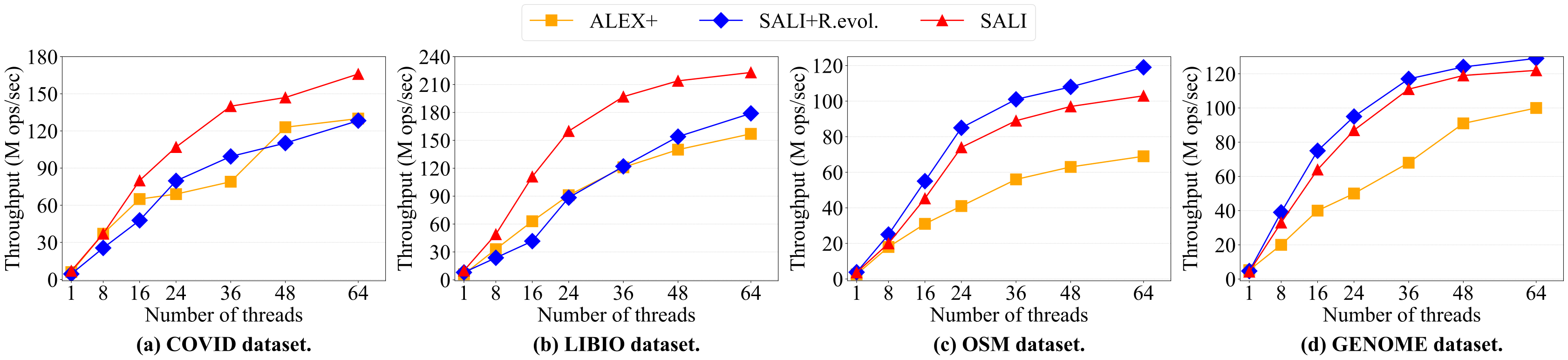}
  \vspace{-1.5em}
  \caption{Performance of read-evolving strategy under read-only workload (uniform distribution).}
  \vspace{-1em}
  \label{fig:figure 25}
\end{figure*}

\section{Analysis of the limitations of \idxname}

\subsection{The limitation of read-evolving strategy}

Insight 4 in Section \ref{Evolving triggered by hot read} states that flattened tree structures under easy datasets do not evolve. 
In such cases, enabling the read-evolving strategy would introduce additional overhead to determine if a target node is a hot node, which would compromise the read performance of the \idxname.

Specifically, we analyze the overhead through experimental evaluation.
When SALI activates read-evolving, the overhead during lookups on a read-only workload (uniform distribution) with different datasets is illustrated in \autoref{fig:figure 24}.
There are three types of overhead during lookups: 
1) the if statement to determine whether the lookup triggers probability calculation, where each thread calculates the probability every 10 lookups, 
2) probability calculation, and 
3) determining whether a node is a hot node. 
Note that, for better clarity in illustrating the performance bottlenecks, all nodes are non-evolving (manually controlled) as node evolving may not always be applicable due to the small tree depth of certain datasets.

In \autoref{fig:figure 24}, $\idxname+Prob.$ represents the throughput of \idxname with the first and second types of overhead, while $\idxname+R.evol.$ represents the throughput of \idxname with all three types of overhead. $\idxname$ represents the read-evolving switch that is turned off.
It can be observed that the first two types of overhead are negligible, and the main overhead comes from the third type.

Therefore, when enabling the read-evolving strategy in easy datasets, the internal nodes of SALI will not undergo evolution, and there will be an additional overhead in determining whether a node is a hot node, potentially leading to a loss of lookup performance.
We acknowledge that evaluating the benefits of enabling read-evolving in specific data distributions can be challenging, and this will be a focal point of our future work.

\autoref{fig:figure 25} illustrates the throughput of the index when read-evolving is enabled under a read-only workload (uniform distribution). 
In contrast to \autoref{fig:figure 24}, the $\idxname+R.evol.$ configuration includes node evolving in the OSM and GENOME datasets, while the experimental results for the COVID and LIBIO datasets align with those in \autoref{fig:figure 24}.
This is because the flattened tree structures in easy datasets do not undergo evolution (refer to Insight 4 in Section \ref{Evolving triggered by hot read}).

\subsection{The limitation of supporting duplicate data}

Currently, \idxname does not support duplicate keys. 
The reason for this is that SALI, being based on the LIPP structure, does not currently support the insertion of duplicate data. 
However, Wu et al.~\cite{wu2021updatable} have suggested that it is relatively straightforward for indexes to accommodate duplicate keys, such as by maintaining a pointer to an overflow list. As part of our future work, we plan to focus on implementing the insertion of duplicate data.

\section{Buffer-based and model-based insertion strategy}

In this section, we will provide a comprehensive discussion of the strengths and weaknesses of the buffer-based insert strategy (out-of-place insert) and the model-based insert strategy (in-place insert). Our aim is to emphasize the significance of choosing different index structures based on specific scenarios.

\subsection{buffer-based insertion strategy}

The major performance limitation of the buffer-based insertion strategy lies in its significant lookup error. 
Unlike the model-based strategy, it lacks the capability to adjust the distribution of stored keys through reserved gaps, which could result in a linearized index structure and reduce or eliminate errors (i.e., ALEX, LIPP). 
The larger lookup error adversely affects the lookup and insertion performance of the index.

From a more detailed perspective, there are two ways to insert data in the buffer.
The first approach involves maintaining the insertion order. 
While this method is more efficient for searching, it requires moving many keys to maintain the order during insertion. 
As a result, it degrades the insertion performance of the buffer. The second approach is the append-only insertion method, where there is no need to maintain the insertion order. 
This method is efficient for write operations but introduces additional overhead during searching and retraining processes. 
Regardless of the approach chosen, both searching and inserting operations in the buffer result in significant time overhead, leading to suboptimal index performance.

However, the buffer-based insertion strategy is not without its opportunities. It demonstrates significant advantages in the following two scenarios:
1) PGM, a type of learned index similar to LSM-Tree, exhibits remarkable performance benefits in write-only workloads. In such scenarios, the buffer-based insertion strategy proves highly effective.
2) In certain environments with strict space limitations, such as embedded systems, the buffer-based insertion strategy's corresponding compact storage (i.e., without gaps) offers space savings of two orders of magnitude compared to B+ trees. Such learned indexes (e.g., PGM) are more suitable for these application scenarios.

In summary, while the buffer-based insertion strategy may have limitations, it demonstrates substantial advantages in specific scenarios, such as write-only workloads and environments with stringent space constraints.

\subsection{model-based insertion strategy}

The model-based insertion strategy offers superior performance in learned indexes due to the following reasons:

1)This strategy introduces gaps between stored keys, altering the cumulative distribution of the stored keys to approximate a linear distribution. 
As a result, the index structure can be approximated as a linear model with minimal error (i.e., ALEX). 
Additionally, techniques like chaining in designs such as LIPP enable conflict resolution, ensuring precise lookups and significantly improving query performance.
2)Unlike the buffer-based insertion strategy, the model-based approach reserves gaps that allow for the efficient insertion of new keys without the need to 'shift' many existing keys.
This substantially enhances the insertion performance of the index.

In summary, the model-based insertion strategy in learned indexes delivers improved performance through CDF optimization, precise lookups, and efficient insertions facilitated by reserved gaps.
Therefore, in the application scenarios of this paper, we have opted to evaluate SALI using the model-based insertion strategy.

\end{document}